\documentclass[11pt]{article}

\usepackage{amsfonts,amssymb,chet,dsfont,graphicx,mathrsfs,pgfplots,subcaption,tikz}

\title{Consistency of Scalar Potentials\\\vspace{0.1cm}from Quantum de Sitter Space}

\author{Jos\'e R.~Espinosa$^{\ast,\dagger,}$\email{espinosa@ifae.es}, Jean-Fran\c{c}ois Fortin$^{\$,}$\email{jean-francois.fortin@phy.ulaval.ca} and Maxime Tr\'epanier$^{\$,}$\email{maxime.trepanier.1@ulaval.ca}}

\affiliation{
$^\ast$Institut de F\'isica d'Altes Energies (IFAE), The Barcelona Institute of Science and Technology (BIST), Campus UAB, E-08193, Bellaterra (Barcelona), Spain\\
$^\dagger$ICREA, Pg. Llu\'is Companys 23, 08010 Barcelona, Spain\\
$^\$$D\'epartement de Physique, de G\'enie Physique et d'Optique,\\Universit\'e Laval, Qu\'ebec, QC G1V 0A6, Canada
}

\abstract{Consistency of the unconventional view of de Sitter space as a quantum theory of gravity with a finite number of degrees of freedom requires that Coleman-De Luccia tunneling rates to vacua with negative cosmological constant should be interpreted as recurrences to low-entropy states.  This demand translates into two constraints, or consistency conditions, on the scalar potential that are generically as follows: 1) the distance in field space between the de Sitter vacuum and any other vacuum with negative cosmological constant must be of the order of the reduced Planck mass or larger and 2) the fourth root of the vacuum energy density of the de Sitter vacuum must be smaller than the fourth root of the typical scale of the scalar potential.  These consistency conditions shed a different light on both outstanding hierarchy problems of the Standard Model of particle physics: the scale of electroweak symmetry breaking and the scale of the cosmological constant.  Beyond the unconventional interpretation of quantum de Sitter space, we complete the analytic understanding of the thin-wall approximation of Coleman-De Luccia tunneling, extend its numerical analysis to generic potentials and discuss the role of gravity in stabilizing the Standard Model potential.}

\date{August 2015} 

\begin{document}

\maketitle



\section{Introduction}\label{SIntro}

Fundamental scalars and the cosmological constant in quantum field theory (QFT) are peculiar beasts.  For fundamental scalars, the symmetry structure of their mass terms in the Lagrangian density is analogous to the structure of their kinetic terms.  For the cosmological constant, the vacuum energy density in the Lagrangian density is simply a constant.  Hence, no symmetries ensure their stability: small fundamental scalar mass terms and a small cosmological constant can suffer from large quantum corrections, the so-called hierarchy problems \cite{Weinberg:1975gm,Weinberg:1979bn,Gildener:1976ai,Susskind:1978ms} \cite{Nernst:1916aa}.  Because of renormalization, the masses of fundamental scalars and the vacuum energy density get additively corrected by the largest scale present, which corresponds to the Planck scale when gravity is considered.  
Obtaining small fundamental scalar mass terms and a small cosmological constant thus necessitates a fine-tuning of the parameters.  Hence, a natural interpretation of small fundamental scalar mass terms and of a small cosmological constant is usually difficult to achieve unless a symmetry or another as-yet-unknown principle solves the issue.

Supersymmetry (SUSY) \cite{Volkov:1973ix,Wess:1974tw} offers such a solution by tying the quantum corrections of fundamental scalar mass terms to their fermionic superpartners, which exhibit naturalness properties \cite{Dimopoulos:1981au,Witten:1981nf,Dine:1981za,Dimopoulos:1981zb,Sakai:1981gr,Kaul:1981hi}, and by providing, for each particle, a superpartner with opposite statistics and the same mass, which leads to a cancellation of the vacuum energy density \cite{Iliopoulos:1974zv} (the situation is different in supergravity).  Unfortunately, supersymmetry must be spontaneously broken \cite{Fayet:1974jb,Fayet:1975ki,O'Raifeartaigh:1975pr} in our universe [which is, as far as we know, very well described by the Standard Model (SM) of particles] and present experimental results on superpartner searches \cite{Aad:2012fqa,Chatrchyan:2012jx} imply some degree of tuning that is usually considered large \cite{Essig:2007kh,Craig:2013cxa} (although what large means for most fine-tuning measures is somewhat subjective \cite{Ellis:1986yg,Barbieri:1987fn,Baer:2014ica,Casas:2014eca}).

Another important consequence of quantum effects is the possible occurrence of tunneling events.  Indeed, for QFT without gravity, only the global minimum of a theory with a potential exhibiting several minima is stable; the remaining minima are unstable due to barrier penetration \cite{Coleman:1977py,Callan:1977pt}.  However, once gravity is included, it is possible for local minima with vanishing or negative cosmological constants to be stable \cite{Coleman:1980aw,Parke:1982pm}, \textit{i.e.} gravity can stabilize them.  When the global minimum has a negative cosmological constant, the conventional view is that all local minima that are not stabilized by gravity eventually decay to the global minimum.  

On the other hand, among the efforts to make sense of quantum gravity in de Sitter (dS) space (a notoriously difficult problem), there is the unconventional possibility that dS space might have a finite number of degrees of freedom \cite{Banks:2000fe,Fischler:2000aa,Witten:2001kn,Goheer:2002vf}.  In the context of that fundamental assumption, the role of minima with negative cosmological constant that might exist besides the dS minimum of interest could be problematic: decays to spaces with negative cosmological constants always correspond to transitions to big crunches \cite{Coleman:1980aw} with the system having an infinite number of degrees of freedom, in contradiction with the assumption of finiteness of the Hilbert space \cite{Banks:2000fe}.  Such tunneling decays to anti-de Sitter (AdS) minima should therefore be forbidden somehow, either by banishing them from the potential altogether or by reinterpreting the tunneling events in some novel way, \textit{e.g.} 
by interpreting them entropically \cite{Aguirre:2006ap}.  Indeed, it was shown numerically in \cite{Aguirre:2006ap} that a local dS minimum in a scalar potential without minima with vanishing cosmological constant
but with minima with negative cosmological constant can be considered stable when the scalar potential satisfies one specific condition.  In such cases, the tunneling event can be consistently interpreted as a recurrence to low-entropy states instead of a full-fledged tunneling event.

In this paper we follow this unconventional point of view  (quantum gravity in dS space with a finite number of degrees of freedom where the notion of transitions to big crunches is forgotten altogether) and derive two constraints on the scalar potential that allow for such reinterpretation of tunneling to AdS events as recurrences to low-entropy states \cite{Aguirre:2006ap}. These conditions are as follows: (1) the distance in field space between the dS vacuum and any other vacuum with negative cosmological constant must be of the order of the reduced Planck mass or larger, and (2) the fourth root of the vacuum energy density of the dS vacuum must be smaller than the fourth root of the typical scale of the scalar potential.

We further speculate about the interplay between such consistency constraints and the usual naturalness criteria and whether the former should replace the latter. Evidently, only a full understanding of quantum gravity in dS space could answer such a question.  However, since the possibility that the hierarchy problems are solved by some unknown mechanism taking place at the Planck scale is not ruled out, in this work we explore the possibility that the problems of fundamental scalars and the cosmological constant in QFT might be red herrings originating from our poor understanding of quantum gravity in dS space.  We then assume that, as long as the two consistency conditions described above are satisfied, there are no hierarchy problems in QFT, although usual na\"ive QFT computations suggest so.

The first consistency condition introduced here, when applied to the SM, can be translated into a lower bound on the Higgs mass $M_h$ with the help of the renormalization group (RG) flow.  The exact lower bound can be calculated from the effective action and is $M_h\gtrsim129.4\,\text{GeV}$ (for a top mass $M_t=173.34\,\text{GeV}$ \cite{ATLAS:2014wva}), which is slightly above the observed Higgs mass $M_h=125.09\pm0.21\,\text{(stat.)}\pm0.11\,\text{(syst.)}\,\text{GeV}$ \cite{Aad:2012tfa,Chatrchyan:2012ufa,Aad:2013wqa,Chatrchyan:2013mxa,Aad:2015zhl,Agashe:2014kda} and might imply new physics below the Planck scale.  Being a lower bound, the first consistency condition is unfortunately not relevant to the Higgs naturalness problem in the framework of the SM.  Another mechanism, \textit{e.g.} SUSY, is thus necessary.  We have nothing more to add on the Higgs naturalness problem.

When applied to the SM, the second consistency condition leads to an upper bound on the vacuum energy density $\rho_\text{vac}$ given by $\rho_\text{vac}^{1/4}\lesssim4.7\times10^{16}\,\text{GeV}$.  This upper bound is in accord with the observed value $\rho_\text{vac}^{1/4}=2.24\pm0.01\,\text{meV}$ \cite{Agashe:2014kda} although it is not as striking as the previous bound.  However, including other fundamental scalars, like axions \cite{Wilczek:1977pj,Weinberg:1977ma,Kim:1979if,Shifman:1979if,Zhitnitsky:1980tq,Dine:1981rt}, leads to somewhat better upper bounds on the vacuum energy density for extensions of the SM.  More importantly, being an upper bound, the second consistency condition is directly relevant to the naturalness problem associated to the cosmological constant.

This paper is organized as follows: Section \ref{SCDL} reviews Coleman-De Luccia (CDL) instantons, focusing on the properties most important for our analysis, specifically the bounce and the background action.  Section \ref{SConsistency} studies the ratio of the instanton action to the background action. This is the crucial quantity that 
governs whether CDL instantons of the dS space can be interpretable as recurrences to low-entropy states (which we dub sub-Poincar\'e recurrences). From a condition on this ratio we then extract the constraints on the scalar potential.  An analytical answer is first found for this ratio in the thin-wall approximation.  In doing so, we complete the analysis of the thin-wall approximation begun in \cite{Coleman:1980aw,Parke:1982pm,Weinberg:2012pjx}.  A numerical analysis of the ratio of the instanton action to the background action is then performed, showing that its generic dependence on two important parameters, the distance between vacua in field space and the vacuum energy density of the false vacuum, stays close to the analytical answer obtained in the thin-wall approximation.  Finally, before concluding in section \ref{SConclusion}, we discuss in section \ref{SSM} the consistency conditions in the context of the SM and some of its extensions, deriving bounds on the SM Higgs mass as a function of the top mass.

Although the view advocated in this paper departs from the conventional one, we note that our results on the analytic understanding in the thin-wall approximation of the numerical conclusion of \cite{Aguirre:2006ap}, its generalization to generic scalar potentials (especially the functional form of the bounce), and the numerical investigation of the role of gravity in the stabilization of the SM potential, are of value on their own.


\section{Coleman-De Luccia Vacuum Decay}\label{SCDL}

This section reviews the work of Coleman and collaborators \cite{Coleman:1977py,Callan:1977pt,Coleman:1980aw} on the tunneling probability per unit time per unit volume for a QFT in a false vacuum following the approach of \cite{Aguirre:2006ap}.  The emphasis is on the bounce and background action that play an important part in our analysis.

\subsection{Bounce and Background Action}\label{SsBounce}

For a QFT with a scalar field $\phi(x)$ in a potential $V(\phi)$ that is assumed to have two minima, only the minimum with the lowest energy density is a stable true vacuum.  The minimum with the highest energy density is an unstable false vacuum.  Indeed quantum tunneling from the false vacuum to the true vacuum is possible through barrier penetration \cite{Coleman:1977py,Callan:1977pt,Coleman:1980aw}.  The decay of the false vacuum proceeds through bubble nucleation, \textit{i.e.} a bubble of true vacuum appears within the false vacuum and expands.  The quantum tunneling event is described by a tunneling probability per unit time per unit volume $\Gamma/V$ that, in the semi-classical limit, is given by an expansion of the form
\eqn{\Gamma/V=Ae^{-B/\hbar}[1+\mathcal{O}(\hbar)].}[EqnTunnel]
Note that $\hbar$ is shown explicitly only in \EqnTunnel.  In the presence of gravity, the nonrenormalizability of the theory forbids a computation of the coefficient $A$ but the bounce $B$ is simply
\eqn{B=S_E(\phi)-S_E(\phi_F).}[EqnBounce]
Here, $S_E(\phi)$ is the Euclidean action for the instanton interpolating between the true vacuum $\phi_T$ and the false vacuum $\phi_F$ of the scalar potential and $S_E(\phi_F)$ is the background Euclidean action.  For a scalar field $\phi(x)$ in a potential $V(\phi)$, the action in the presence of gravity (using the mostly-minus metric) is
\eqn{S=\int d^4x\,\sqrt{-g}\left[\frac{1}{2}\,g^{\mu\nu}\partial_\mu\phi\,\partial_\nu\phi-V(\phi)-\frac{m_P^2}{2}R\right],}[EqnAction]
where $m_P^2=(8\pi G_N)^{-1}=M_P^2/(8\pi)$ is the reduced Planck mass, $M_P^2=G_N^{-1}$ is the Planck mass and $R$ is the scalar curvature (Ricci scalar).  The instanton is described by a scalar field configuration $\phi(z)$ and a Euclidean manifold $(ds)^2=(dz)^2+\rho(z)^2(d\Omega)^2$, assuming the dominant instanton contribution to \EqnTunnel is $O(4)$-symmetric.

\begin{figure}[!t]
\centering
\resizebox{8cm}{!}{
\includegraphics{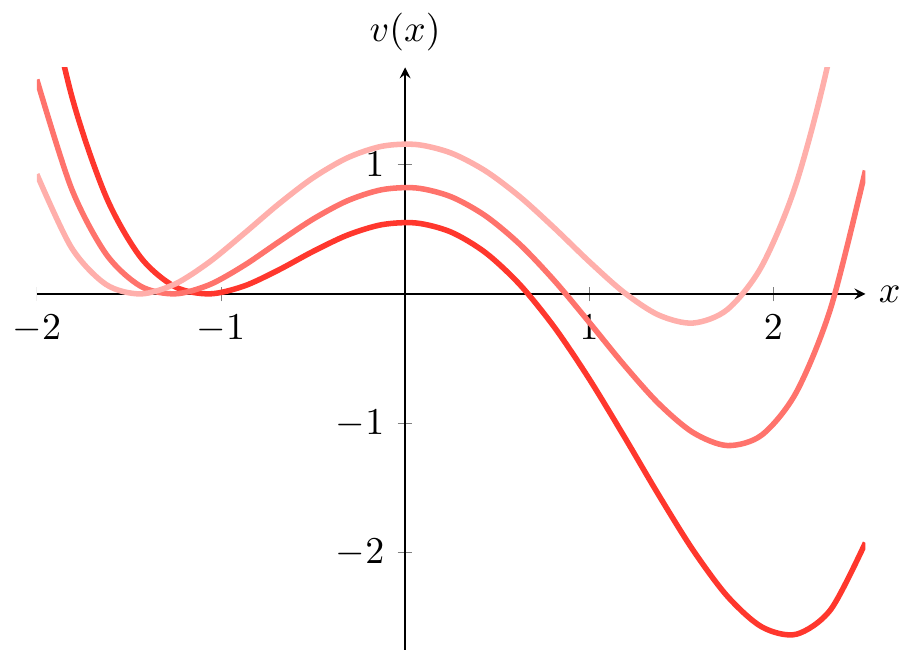}
}
\caption{Scalar potentials with $(a,b)=(1.5,1.0)$ (lower curve), $(a,b)=(1.5,0.5)$ (middle curve), and $(a,b)=(1.5,0.1)$ (upper curve) for $\xi=0$.}
\label{FigScalarPotential}
\end{figure}

Before proceeding further, it is convenient to describe the scalar potential as well as introduce dimensionless quantities.  For our numerical analysis, the scalar potential is chosen to be of the form
\eqna{
V(\phi)&=\mu^4v\left(x=\frac{\phi}{M}\right)=\mu^4[f(x)-(1-\xi)f(x_F)-\xi f(x_T)],\\
f(x)&=\frac{x^4}{4}-\frac{bx^3}{3}-\frac{a^2x^2}{2},
}[EqnPotential]
where $x=\phi/M$ and the false and the true minima are located at $x_F=\phi_F/M$ and $x_T=\phi_T/M$ with $v_F\equiv v(x_F)$ and $v_T\equiv v(x_T)$ respectively.  This effective potential is not well-suited for processes for which the scalar field $\phi$ explores asymptotically large values in units of $M$, \textit{i.e.} when $|x|\gg1$.  It is otherwise sufficient from an effective field theory point of view.  The characteristic scales $M$ and $\mu$ of the scalar potential in \EqnPotential are chosen such that most variations in the dimensionless scalar potential are order one, \textit{e.g.} $|\Delta\phi|\equiv|\phi_F-\phi_T|\approx M$ and $\Delta V\equiv V(\phi_F)-V(\phi_T)\approx\mu^4$.  Three illustrative choices for the scalar potential parameters are $(a,b)=(1.5,1.0)$, which corresponds to a scalar potential with a small energy barrier; $(a,b)=(1.5,0.5)$, which corresponds to a natural scalar potential with all variations of the same order; and $(a,b)=(1.5,0.1)$, which corresponds to a scalar potential with a small inter-vacua energy difference.  The scalar potentials are shown in figure \ref{FigScalarPotential}.  Note finally that the parameter $\xi$ controls the energy density of the true and the false vacua, with $v_T=0$ for $\xi=1$ and $v_F=0$ for $\xi=0$, or more generically $v_T=-(1-\xi)\Delta v$ and $v_F=\xi\Delta v$ with $\Delta v=\Delta V/\mu^4=v_F-v_T$ being the dimensionless vacuum energy difference between the two vacua.  This last definition is a convention adopted for all scalar potentials, not just for the ones used in our numerical analysis \EqnPotential.

Before analyzing the bounce, it is convenient to introduce the dimensionless quantities
\eqn{r=\frac{\mu^2}{M}\rho,\quad\quad s=\frac{\mu^2}{M}z,\quad\quad\epsilon^2=\frac{M^2}{3m_P^2}.}[EqnDimless]
From \EqnAction, the metric, and \EqnDimless, the coupled Euclidean scalar field and Einstein's equations for the dimensionless quantities are
\eqna{
\frac{d^2x}{ds^2}+\frac{3}{r}\frac{dr}{ds}\frac{dx}{ds}&=\frac{dv}{dx},\\
\left(\frac{dr}{ds}\right)^2&=1+\epsilon^2r^2E,
}[EqnEOM]
with $E$ being the Euclidean field energy,
\eqn{E=\frac{1}{2}\left(\frac{dx}{ds}\right)^2-v.}[EqnE]
Note that the dimensionless parameter $\epsilon$ determines the size of the gravitational effects.

Since $r(0)=0$, to avoid any singularities the boundary conditions for a non-singular instanton are
\eqn{x(0)=x_0,\quad\quad\frac{dx}{ds}(0)=0,\quad\quad r(0)=0,\quad\quad\frac{dr}{ds}(0)=1.}
The parameter $x_0$ must be tuned in the neighborhood of $x_T$ for a non-singular solution to be found.  Moreover, $x_0$ is tuned such that $dx/ds$ has only two zeros, \textit{i.e.} the instanton is single-pass.  Hence the derivative of the field, $dx/ds$, vanishes at both ends of the interval.  Thus, for compact instantons the field boundary conditions do not have to equal the vacuum field values.  This is usually interpreted as a statement that the system is thermal.  For non-compact instantons, the field must reach its false vacuum field value; otherwise, the bounce is infinite.  This can also be understood from the asymptotics of the non-compact instanton solutions.  Note that for a large enough $\epsilon$, only the Hawking-Moss (HM) instanton \cite{Hawking:1981fz} exists.

In terms of the dimensionless quantities \EqnDimless the Euclidean action is therefore
\eqna{
S_E&=2\pi^2\left(\frac{M}{\mu}\right)^4\int_{s=0}^{s=s_\text{max}}ds\left\{r^3\left[\frac{1}{2}\left(\frac{dx}{ds}\right)^2+v\right]+\frac{1}{\epsilon^2}\left[r^2\frac{d^2r}{ds^2}+r\left(\frac{dr}{ds}\right)^2-r\right]\right\}\\
&=-\frac{4\pi^2}{\epsilon^2}\left(\frac{M}{\mu}\right)^4\left[\int_{s=0}^{s=s_\text{max}}ds\,r(1-\epsilon^2r^2v)-\frac{r_\text{max}^2}{2}\left.\frac{dr}{ds}\right|_{s=s_\text{max}}\right],
}[EqnActionO]
where the equations of motion \EqnEOM and the Euclidean field energy \EqnE have been used.  Here $s_\text{max}$ is finite [\textit{i.e.} $r(s_\text{max})$ vanishes] for compact instantons and $s_\text{max}$ is infinite [\textit{i.e.} $r(s_\text{max})$ is infinite] for non-compact instantons.  Notice the explicit surface term in the Euclidean action.  Since the instanton solution for $r(s)$ must match the background solution at infinity, the surface term contribution to the bounce always cancels between the instanton Euclidean action and the background Euclidean action.  From \EqnActionO it is now easy to see that $\mu\ll M$ for the semi-classical approximation to be valid.

The equations of motion \EqnEOM can be solved analytically in the background since the field is fixed at the false vacuum value $x=x_F$.  The solutions are
\eqna{
r_\text{dS}(s)&=\tilde{r}\sin(s/\tilde{r}),\\
r_\text{M}(s)&=s,\\
r_\text{AdS}(s)&=\tilde{r}\sinh(s/\tilde{r}),
}[EqnBackgroundEOM]
for dS spaces, Minkowski space, and AdS spaces respectively.  Here, $\tilde{r}=(\epsilon\sqrt{|v_F|})^{-1}$ where $v_F$ is the (dimensionless) false vacuum energy density for dS and AdS spaces.  Since Euclidean dS space is compact while Euclidean Minkowski and AdS spaces are non-compact, one has finite $s_\text{max}$ with vanishing $r(s_\text{max})=0$ for dS spaces and infinite $s_\text{max}$ with infinite $r_\text{max}\equiv r(s_\text{max})$ for Minkowski and AdS spaces.

The background Euclidean action is easily computed from the general formula \EqnActionO with constant field $x=x_F$.  The result is simply
\eqna{
S_\text{dS}&=-\frac{8\pi^2}{3\epsilon^4v_F}\left(\frac{M}{\mu}\right)^4,\\
S_\text{M}&=0,\\
S_\text{AdS}&=-\frac{4\pi^2}{\epsilon^2}\left(\frac{M}{\mu}\right)^4\left\{\frac{1}{3\epsilon^2v_F}\left[1-(1-\epsilon^2r_\text{max}^2v_F)^{3/2}\right]-\frac{r_\text{max}^2}{2}\left.\frac{dr}{ds}\right|_{s=s_\text{max}}\right\},
}[EqnBackground]
where again $r_\text{max}$ is infinite for AdS spaces.  
Notice that the limit of vanishing false vacuum energy density $v_F\rightarrow0$ reproduces the flat space result only for AdS spaces since both AdS and Minkowski spaces are non-compact.  Indeed, for AdS spaces the background Euclidean action goes to zero as $v_F\rightarrow0$ while for dS spaces the background Euclidean action diverges as $v_F\rightarrow0$.

From \EqnActionO and \EqnBackground, the bounce \EqnBounce is then
\eqna{
B^\text{c}&=-\frac{4\pi^2}{\epsilon^2}\left(\frac{M}{\mu}\right)^4\left[\int_{s=0}^{s=s_\text{max}}ds\,r(1-\epsilon^2r^2v)-\frac{2}{3\epsilon^2v_F}\right],\\
B^\text{nc}&=-\frac{4\pi^2}{\epsilon^2}\left(\frac{M}{\mu}\right)^4\left\{\int_{s=0}^{s=s_\text{max}}ds\,r(1-\epsilon^2r^2v)-\frac{1}{3\epsilon^2v_F}\left[1-(1-\epsilon^2r_\text{max}^2v_F)^{3/2}\right]\right\},
}[EqnBounceO]
where the subscripts ${}^\text{c}$ and ${}^\text{nc}$ denote compact ($v_F>0$) and non-compact ($v_F\leq0$) instantons, respectively.  For compact instantons both the $v_F\to0$ and $\epsilon\to0$ limits give wrong results because the limits do not commute with the integrals.  Indeed, it was necessary that $r(s_\text{max})=0$ to compute the background Euclidean action in dS space, which is never the case in the two limits.  As already mentioned, the limit $v_F\to0$ for non-compact instantons is consistent.  Moreover, the $\epsilon\to0$ limit for non-compact instantons gives the expected result when gravity is absent,
%
\eqn{\lim_{\epsilon\to0}B^\text{nc}=2\pi^2\left(\frac{M}{\mu}\right)^4\int_{r=0}^{r=\infty}dr\,r^3\left[\frac{1}{2}\left(\frac{dx}{dr}\right)^2+v-v_F\right]=B^{\epsilon=0}.}

Although $A$ cannot be computed, it is nevertheless possible to estimate its size.  In the absence of gravity, the prefactor $A$ can be calculated by a careful treatment of the modes and gives $A\approx B^{n/2}\det(\cdot)$ \cite{Callan:1977pt} where $n$ is the number of collective coordinates ($n=4$ for the spacetime position of the bubble) and the determinant is over the remaining modes (with non-vanishing eigenvalues).  Since $\det(\cdot)\approx M^4$ the prefactor $A\approx M^4(M/\mu)^8$.  Finally, putting \EqnTunnel and \EqnBounceO together, one obtains
\eqn{\Gamma/V\approx M^4\left(\frac{M}{\mu}\right)^8e^{-B},}[EqnGamma]
for the tunneling probability per unit time per unit volume induced by quantum fluctuations.


\section{Consistency Conditions for Quantum Field Theory in dS Space}\label{SConsistency}

In this section we review the argument leading to the assumption of finiteness of the Hilbert space of a dS space \cite{Banks:2000fe,Fischler:2000aa,Witten:2001kn,Goheer:2002vf} and introduce the constraints on the scalar potential this assumption implies.  We then argue that the constraints obtained should be understood as consistency conditions that supersede the usual QFT hierarchy problems.

\subsection{dS Space with a Finite Number of Degrees of Freedom}\label{SsQG}

Since it is possible to interpret the Euclidean dS path integral as a thermal ensemble, dS space, which in static coordinates is described by the metric
\eqn{(ds)^2=\left(1-\frac{R^2}{R_\text{dS}^2}\right)(dt)^2-\left(1-\frac{R^2}{R_\text{dS}^2}\right)^{-1}(dR)^2-R^2(d\theta)^2-R^2\sin^2\theta(d\phi)^2,}
where $R_\text{dS}=\sqrt{3/\Lambda_\text{dS}}$ is the cosmological horizon (which was denoted by $\frac{M}{\mu^2}\tilde{r}$ in subsection \ref{SsBounce}) and $\Lambda_\text{dS}$ is the cosmological constant, has a corresponding temperature $T_\text{dS}=(2\pi R_\text{dS})^{-1}$ \cite{Figari:1975km} and an associated entropy $\mathcal{S}_\text{dS}$ \cite{Gibbons:1977mu}.  As for the Bekenstein-Hawking entropy of a black hole, the dS entropy is
\eqn{\mathcal{S}_\text{dS}=\frac{A}{4G_N}=-S_\text{dS},}[EqnEntropy]
where the last identity comes from the definition of the cosmological constant $\Lambda_\text{dS}=V_F/m_P^2$.  Interpreting the dS entropy as any other entropy, \textit{i.e.} as the logarithm of the number of quantum states of the dS space, leads to the conclusion that the Hilbert space $\mathcal{H}$ describing the quantum theory on dS space is finite \cite{Banks:2000fe}.  Indeed, since a finite entropy indicates that the total number of quantum states necessary to describe the physics is finite, it is possible to consider the relevant Hilbert space as the space built from those quantum states only, hence a finite-dimensional Hilbert space.  The dimension $\dim\mathcal{H}$ of the Hilbert space is related to the dS entropy through the usual statistical equation, $\dim\mathcal{H}=\exp(\mathcal{S}_\text{dS})$.

The relation \EqnEntropy between the dS entropy and the dS background Euclidean action \EqnBackground suggests an entropic interpretation of CDL quantum tunneling from a false dS vacuum to any vacua with negative cosmological constants.  Indeed, in the usual CDL quantum tunneling picture, the spacetime inside the bubble (where the scalar field lingers close to the true vacuum), obtained from analytic continuation, is an expanding-and-contracting open Friedmann-Lema\^itre-Robertson-Walker (FLRW) spacetime, leading to a big crunch \cite{Coleman:1980aw}.  Thus, the fate of the false dS vacuum, an open FLRW universe with an obviously infinite number of quantum states, is also at odds with the finiteness of the Hilbert space and again suggests an entropic point of view for the CDL tunneling rate.  Moreover, in transitioning from the false dS vacuum to a vacuum with negative vacuum energy density, the scalar field does not remain close to the true AdS vacuum, instead it probes the entire scalar potential until the energy density approaches the Planck scale and the effective field theory approximation is badly broken \cite{Coleman:1980aw}.  Quantum tunneling does not let the scalar field settle in the true AdS vacuum.

An obvious proposition alleviating these issues is to demand that the tunneling probability per unit time per unit volume for CDL instantons out of a ``false'' dS vacuum is understandable as a sub-Poincar\'e recurrence, \textit{i.e.} a recurrence to a low-entropy state,\footnote{The unlikely occurrence of all the gas particles gathering in one small corner of a closed volume would be a statistical equivalent.} such that the assumption of a finite-dimensional Hilbert space for quantum dS space stays consistent \cite{Aguirre:2006ap}. A CDL quantum tunneling event would simply be a sub-Poincar\'e recurrence where a state resembling the AdS vacuum appears spontaneously in some region of the dS spacetime.  This is possible when $B^\text{c}\approx\mathcal{S}_\text{dS}$, \textit{i.e.} when the CDL bounce is of the same order as (but smaller than) the dS entropy.

\subsection{Thin-Wall Approximation}\label{SsThinWall}

As argued, the tunneling probability per unit time per unit volume \EqnGamma can be interpreted as a sub-Poincar\'e recurrence when $B^\text{c}\approx\mathcal{S}_\text{dS}$, allowing an entropic understanding of quantum dS space.  From \EqnBounceO and \EqnEntropy, the compact CDL bounce can be rewritten as
\eqn{B^\text{c}=\mathcal{S}_\text{dS}(1-R),}[EqnBR]
where the ratio of the instanton action to the background action is
\eqn{R=1-\frac{B^\text{c}}{\mathcal{S}_\text{dS}}=\frac{S_E(\phi)}{S_E(\phi_F)}=\frac{3\epsilon^2v_F}{2}\int_{s=0}^{s=s_\text{max}}ds\,r(1-\epsilon^2r^2v).}[EqnR]
To satisfy the condition $B^\text{c}\approx\mathcal{S}_\text{dS}$, it is clear from \EqnBR that the ratio $R$ must be small.  To understand how $R$ can be small, its behavior as a function of $\epsilon$, $\xi$, and the parameters of the scalar potential (here, $a$ and $b$ in our numerical analysis) must be analyzed.  It is fortunately possible to investigate the analytical properties of the ratio $R$ in the thin-wall approximation.  To proceed, it is necessary to extend and complete the analysis of \cite{Coleman:1980aw,Parke:1982pm,Weinberg:2012pjx}.

The thin-wall approximation is relevant for scalar potentials with small vacuum energy differences between the false vacuum and the true vacuum when compared to the typical scale of the scalar potentials, \textit{i.e.} $\Delta v\ll1$.  In that regime, the scalar potential can be written as
\eqn{v(x)=v_0(x)+v_1(x)\Delta v,}
where $v_0(x)$ is a function with minima at $x_F$ and $x_T$ such that $v_0(x_F)=v_0(x_T)=v_F$ and $v_1(x)$ is a function such that $v_1(x_F)=0$ while $v_1(x_T)=-1$.  In this approximation, the scalar field equation of motion \EqnEOM simplifies to
\eqn{\frac{d^2x}{ds^2}=\frac{dv_0}{dx},}[EqnEOMa]
where the term proportional to $dx/ds$ is discarded since it is negligible as is shown below.  The solution to \EqnEOMa is given by
\eqn{\left(\frac{dx}{ds}\right)^2=2[v_0(x)-v_0(x_T)],}[EqnEOMsoln]
since $x(0)=x_T$.  Combined with \EqnE, this leads to $E=-[v_F+v_1(x)\Delta v]$.  It also implies that $x(s)$ interpolates between $x_T$ at $s=0$ and $x_F$ at large $s$.  The solution to \EqnEOMsoln leads to
\eqn{\bar{s}-s=\int_{(x_F+x_T)/2}^xdx\,\{2[v_0(x)-v_0(x_T)]\}^{-1/2},}
which gives $x(s)$ in terms of the integration constant $\bar{s}$ defined as the coordinate at which the field $x(\bar{s})=(x_F+x_T)/2$.\footnote{The negative branch of the square root has been chosen here since the false and the true vacua are assumed such that $x_F<x_T$.}  Being a coordinate, $\bar{s}$ has no physical meaning; however, $\bar{r}=r(\bar{s})$ does and corresponds to the curvature radius of the wall dividing the true vacuum inside the bubble from the false vacuum outside the bubble.  Solving \EqnEOM gives $r(s)$ in terms of the integration constant $\bar{r}$ that still needs to be computed.

To determine $\bar{r}$, it is convenient to assume that it is large with respect to the wall thickness, \textit{i.e.} $\bar{r}\gg1$, leading to three distinct regions (inside the bubble, the wall of the bubble, and outside the bubble).  This assumption is justified together with the assumption that the term proportional to $dx/ds$ in \EqnEOM is negligible.  Then $\bar{r}$ is obtained by demanding that the bounce is stationary under variations in $\bar{r}$.

Since the field inside the bubble loiters near the true vacuum, $dx/ds=0$ and from \EqnEOM $ds=dr(1-\epsilon^2r^2v_T)^{-1/2}$.  Thus from \EqnActionO the bounce inside the bubble is given by
\eqna{
B_\text{inside}&=-\frac{4\pi^2}{\epsilon^2}\left(\frac{M}{\mu}\right)^4\int_0^{\bar{r}}dr\,\left[r(1-\epsilon^2r^2v_T)^{1/2}-r(1-\epsilon^2r^2v_F)^{1/2}\right]\\
&=-\frac{4\pi^2}{\epsilon^2}\left(\frac{M}{\mu}\right)^4\left\{\frac{1}{3\epsilon^2v_T}\left[1-(1-\epsilon^2\bar{r}^2v_T)^{3/2}\right]-\frac{1}{3\epsilon^2v_F}\left[1-(1-\epsilon^2\bar{r}^2v_F)^{3/2}\right]\right\}.
}[EqnBinside]

In the wall, $r$ can be approximated by $\bar{r}$ since the wall is thin, which gives the following result for the bounce,
\eqn{B_\text{wall}=4\pi^2\left(\frac{M}{\mu}\right)^4\bar{r}^3\int ds\,[v(x)-v_F]=2\pi^2\left(\frac{M}{\mu}\right)^4\bar{r}^3\bar{a},}[EqnBwall]
where, with the help of \EqnEOMsoln,
\eqna{
\bar{a}&=\int ds\,2[v(x)-v_F]\approx\int ds\,2[v_0(x)-v_0(x_F)]\\
&=\int_{x_F}^{x_T}dx\,\{2[v_0(x)-v_0(x_T)]\}^{1/2}\approx\int_{x_F}^{x_T}dx\,\{2[v(x)-v_T]\}^{1/2}.
}[Eqna]
For the one-dimensional dimensionless action $\bar{a}$ in \Eqna, any result is equivalent up to negligible $\mathcal{O}(\Delta v)$ corrections.  However, the last equality is the most convenient when extending the results to more generic potentials.

Finally, outside the bubble the field remains close to the false vacuum with $dx/ds=0$, which na\"ively implies that the bounce vanishes.  This is, however, true only when $\epsilon$ is small, \textit{i.e.} $\epsilon\leq\epsilon_c$ where
\eqn{\epsilon_c=\frac{2\sqrt{\Delta v}}{3\bar{a}},}[Eqnecsoln]
as is computed shortly.  In fact, there are effectively three interesting regimes corresponding to $\epsilon\leq\epsilon_c$, $\epsilon\gtrsim\epsilon_c$, and $\epsilon\gg\epsilon_c$.

\begin{figure}[!t]
\centering
\resizebox{16cm}{!}{
\includegraphics{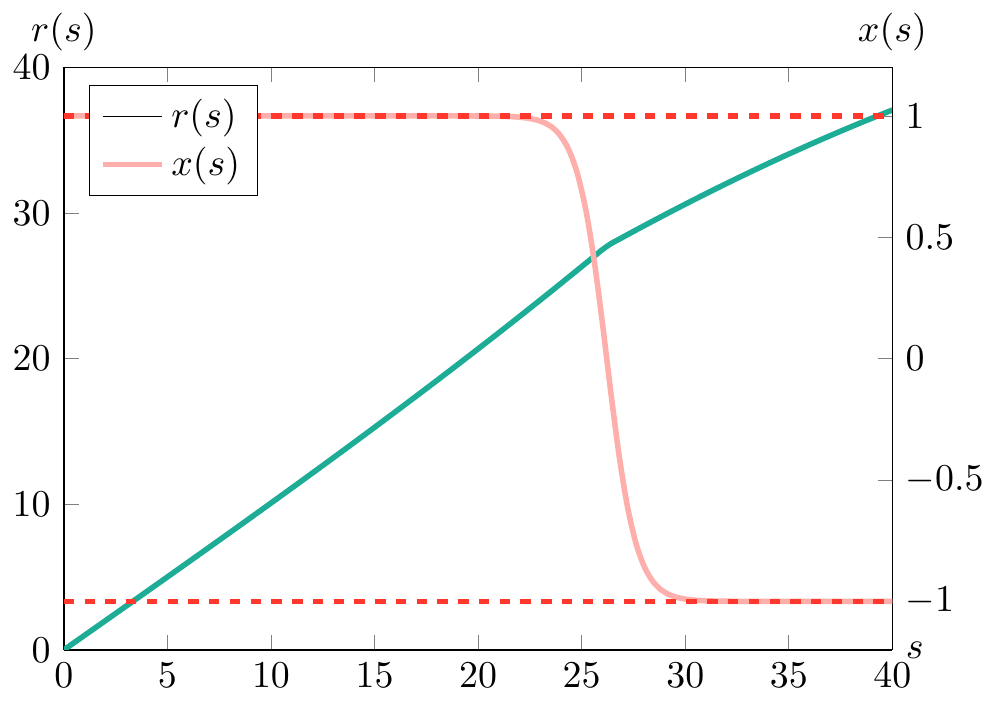}
\hspace{2cm}
\includegraphics{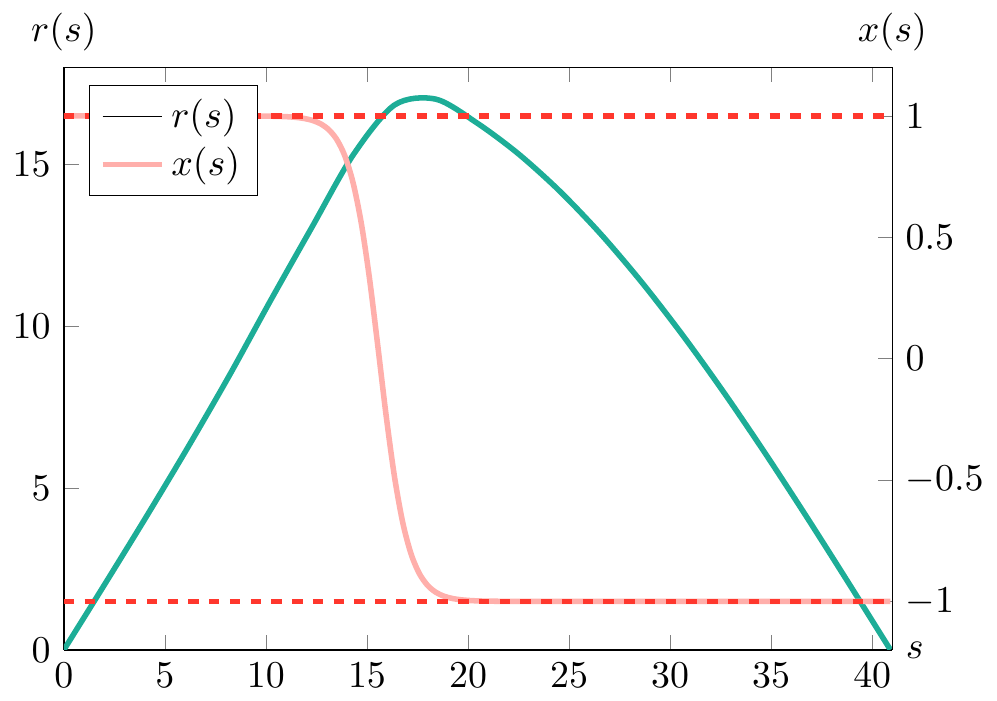}
\hspace{2cm}
\includegraphics{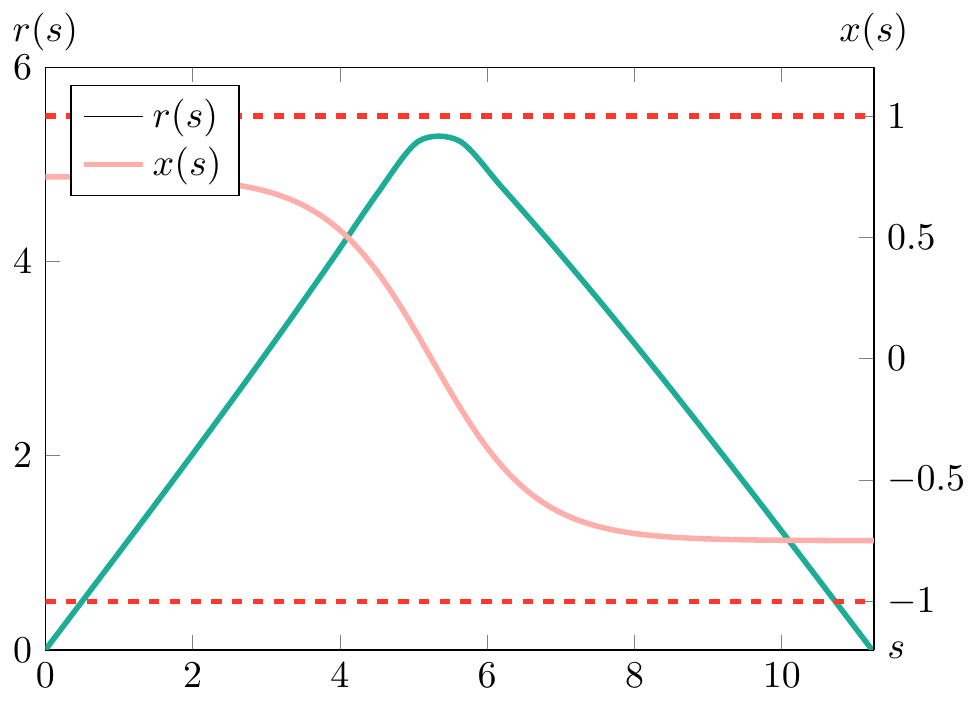}
}
\caption{Qualitative behavior of instanton solutions in the thin-wall approximation for $\epsilon\leq\epsilon_c$ (left), $\epsilon\gtrsim\epsilon_c$ (middle), and $\epsilon\gg\epsilon_c$ (right).}
\label{FigInstanton}
\end{figure}

\subsubsection{The $\epsilon\leq\epsilon_c$ Regime}

In this first regime where $\epsilon\leq\epsilon_c$ the na\"ive result is correct, $ds=dr(1-\epsilon^2r^2v_F)^{-1/2}$, and one has
\eqn{B_\text{outside}^{\epsilon\leq\epsilon_c}=0,}[EqnBoutsidees]
which leads to
\eqna{
\bar{r}^{\epsilon\leq\epsilon_c}&=\frac{4}{3\bar{a}[(\epsilon^2-\epsilon_c^2)^2+4\epsilon_c^2\epsilon^2\xi]^{1/2}},\\
\bar{B}^{\epsilon\leq\epsilon_c}&=-\frac{4\pi^2}{3\epsilon^4\xi\Delta v}\left(\frac{M}{\mu}\right)^4\frac{\epsilon^2(2\xi-1)+\epsilon_c^2-[(\epsilon^2-\epsilon_c^2)^2+4\epsilon_c^2\epsilon^2\xi]^{1/2}}{(1-\xi)[(\epsilon^2-\epsilon_c^2)^2+4\epsilon_c^2\epsilon^2\xi]^{1/2}},
}[EqnrBes]
after demanding that the total bounce $B_\text{inside}+B_\text{wall}+B_\text{outside}^{\epsilon\leq\epsilon_c}$ is stationary with respect to $\bar{r}$.  This is the solution found in \cite{Coleman:1980aw,Parke:1982pm,Weinberg:2012pjx} and it exists only for $\epsilon\leq\epsilon_c$ as given by \Eqnecsoln.  The qualitative behavior of such solutions is shown in the left panel of figure \ref{FigInstanton}.  The important feature to observe is the sign of $dr/ds$, which stays positive through the wall, as can be seen from \EqnEOM, making the bounce outside the bubble vanish.

Note that \EqnrBes is also correct when $\xi\leq0$ and $\xi>1$.  Moreover, for transitions from and to minima with non-positive vacuum energy densities (\textit{i.e.} with $\xi\leq0$) the instanton solution exists only as long as the square root in \EqnrBes is real.  As described in \cite{Coleman:1980aw}, there are no instanton solutions when the square root is imaginary and the corresponding Minkowski or AdS space is stable.

\subsubsection{The $\epsilon>\epsilon_c$ Regime}

But what about $\epsilon>\epsilon_c$ ? Considering $\epsilon_c$ is not in any way special, surely there must exist instanton solutions in that regime.  In fact, instanton solutions do exist but the bounce outside the bubble does not vanish like in \EqnBoutsidees.  Indeed, when $\epsilon>\epsilon_c$, the second term in the equation of motion for $r(s)$ in \EqnEOM becomes of order $1$ in the wall at which point $r(s)$ reaches a maximum and $dr/ds$ changes sign and becomes negative, implying $ds=-dr(1-\epsilon^2r^2v_F)^{-1/2}$ with the all important minus sign for the instanton solution.\footnote{When this work was completed we realized that the all important minus sign is already mentioned in \cite{Weinberg:2012pjx}.}  This can be understood from \EqnEOM and \EqnrBes right outside the wall at $r=\bar{r}$ when $\epsilon=\epsilon_c$, which gives $dr/ds=0$.  Hence, the bounce is given by
\eqn{B_\text{outside}^{\epsilon>\epsilon_c}=-\frac{8\pi^2}{\epsilon^2}\left(\frac{M}{\mu}\right)^4\int_{\bar{r}}^0dr\,r(1-\epsilon^2r^2v_F)^{1/2}=\frac{8\pi^2}{\epsilon^2}\left(\frac{M}{\mu}\right)^4\frac{1}{3\epsilon^2v_F}\left[1-(1-\epsilon^2\bar{r}^2v_F)^{3/2}\right].}[EqnBoutsideel]
Although the instanton solution and the background solution agree at infinity, \EqnBoutsidees and \EqnBoutsideel are different and the bounce does not vanish outside the bubble when $\epsilon>\epsilon_c$.  Combining $B_\text{inside}+B_\text{wall}+B_\text{outside}^{\epsilon>\epsilon_c}$ and enforcing that the result is stationary with respect to variations in $\bar{r}$ leads to
\eqna{
\bar{r}^{\epsilon>\epsilon_c}&=\frac{4}{3\bar{a}[(\epsilon^2-\epsilon_c^2)^2+4\epsilon_c^2\epsilon^2\xi]^{1/2}},\\
\bar{B}^{\epsilon>\epsilon_c}&=-\frac{4\pi^2}{3\epsilon^4\xi\Delta v}\left(\frac{M}{\mu}\right)^4\frac{\epsilon^2(2\xi-1)+\epsilon_c^2-[(\epsilon^2-\epsilon_c^2)^2+4\epsilon_c^2\epsilon^2\xi]^{1/2}}{(1-\xi)[(\epsilon^2-\epsilon_c^2)^2+4\epsilon_c^2\epsilon^2\xi]^{1/2}}.
}[EqnrBel]
This solution, which was not discussed in \cite{Coleman:1980aw,Parke:1982pm},\footnote{But it is discussed in \cite{Weinberg:2012pjx} as already mentioned.} exists only when $\epsilon>\epsilon_c$.  A qualitative sketch of such solutions can be seen in the middle ($\epsilon\gtrsim\epsilon_c$) and right ($\epsilon\gg\epsilon_c$) panels of figure \ref{FigInstanton}, which shows the important change of sign of $dr/ds$ in the wall leading to a mismatch between the outside instanton action and the outside background action.  The qualitative differences between the middle and right panels of figure \ref{FigInstanton} are assessed soon.

Again, note that \EqnrBel is correct when $\xi>1$.  However, when $\xi\leq0$, \textit{i.e.} for transitions from and to minima with non-positive vacuum energy densities, \EqnrBel is not correct since the instanton solution does not match the background solution at infinity; the bounce is therefore infinite.  Indeed, the negative branch of the square root in \EqnEOM is not consistent with \EqnBackground in those cases.  Thus Minkowski and AdS spaces are stable when $\epsilon>\epsilon_c$ in the thin-wall approximation.

Before distinguishing between the two regimes with $\epsilon>\epsilon_c$, let us first state the obvious: \EqnrBes and \EqnrBel are exactly the same even though the bounces as functions of $\bar{r}$ were not.  Hence, the condition on $\epsilon$ can be forgotten altogether and both regimes are described by the same equations.  As is shown below, this fact does not preclude some type of phase transition in the behavior of the bounce as one moves from one regime to another.

\subsubsection{Regime of Validity}

For the thin-wall approximation to be valid, the term proportional to $dx/ds$ in \EqnEOM must be negligible.  Away from the wall it is negligible since $dx/ds$ is negligible.  At the wall, it is negligible because $(1/r)(dr/ds)$ is small.  This last condition can be investigated with the help of \EqnEOM rewritten as
\eqn{\frac{1}{r^2}\left(\frac{dr}{ds}\right)^2=\frac{1}{r^2}+\epsilon^2E.}
In the thin-wall approximation, \EqnEOMsoln implies the Euclidean field energy is approximately constant in all three regions (inside the bubble, in the wall and outside the bubble) and its absolute value is always smaller than $\Delta v$.  Hence, at the wall $r$ is replaced by $\bar{r}$ and $E$ can be conservatively replaced by $\Delta v$.  Thus, the thin-wall approximation is reliable as long as $\bar{r}$ is large and $\pm\epsilon^2\Delta v$ is small.  Since the thin-wall approximation corresponds by definition to the limit $\Delta v\ll1$, it remains to ensure that $\bar{r}$ and $(\epsilon\sqrt{\Delta v})^{-1}$ are large for the values of $\epsilon$ and $\xi$ that are being investigated, which brings us to the distinction between the regimes $\epsilon\gtrsim\epsilon_c$ and $\epsilon\gg\epsilon_c$.

First, $(\epsilon\sqrt{\Delta v})^{-1}$ is large as long as $\epsilon\ll1/\sqrt{\Delta v}$.  Since $\epsilon_c=\sqrt{\Delta v}$ up to a factor of order $1$, this condition can be rewritten as $\epsilon\ll1/\epsilon_c$ and is always satisfied for $\epsilon\leq\epsilon_c$ in the thin-wall approximation.  Moreover, it can also be satisfied for some $\epsilon>\epsilon_c$ in the thin-wall approximation.  Then, $\bar{r}$ is large as long as $(\epsilon^2-\epsilon_c^2)^2+4\epsilon_c^2\epsilon^2\xi\ll1$.  Again, this condition is always true when $\epsilon\leq\epsilon_c$ in the thin-wall approximation and it can also be verified for some $\epsilon>\epsilon_c$.  Therefore, the thin-wall approximation is justified for $\epsilon$ smaller and larger than $\epsilon_c$ as long as $\epsilon$ is not too large as dictated by the inequalities $\epsilon\ll1/\epsilon_c$ and $(\epsilon^2-\epsilon_c^2)^2+4\epsilon_c^2\epsilon^2\xi\ll1$.  Note that the second inequality is more constraining in the thin-wall approximation since $\epsilon_c\ll1$.  Hence, the regime with $\epsilon\gtrsim\epsilon_c$ corresponds to the case where the inequalities are satisfied and the thin-wall approximation is valid [\textit{i.e.} \EqnrBel is valid] while the regime with $\epsilon\gg\epsilon_c$ corresponds to the case where the inequalities are not satisfied and the thin-wall approximation is unjustified.  For the extreme $\epsilon$ regime, the instanton solutions are thermal, \textit{i.e.} they do not start and end at the true and false vacua, respectively, but instead start and end higher up the scalar potential energy barrier (see the right panel of figure \ref{FigInstanton}).  As $\epsilon$ is increased, the start and end points of the instanton solution converge towards the top of the scalar potential until the solution corresponds to the HM instanton.  Moreover, the field does not loiter significantly around its initial and final values but instead transitions slowly between them.  In other words, $dx/ds$ is not zero for most of the time inside and outside the bubble and the bubble wall is not thin.  Thus, the approximation discussed above is unwarranted; the computation of the bounce breaks down in all three regions (inside the bubble, in the wall, and outside the bubble).

\subsubsection{Results}

All that being said, from \EqnrBes and \EqnrBel the bubble radius and the bounce action in the thin-wall approximation are the same for all $\epsilon$,
\eqna{
\bar{r}&=\frac{2}{\epsilon_c\sqrt{\Delta v}}\frac{1}{[(\alpha-1)^2+4\alpha\xi]^{1/2}},\\
\bar{B}&=\frac{8\pi^2}{3\epsilon_c^4\Delta v}\left(\frac{M}{\mu}\right)^4\frac{\alpha(2\xi-1)+1-[(\alpha-1)^2+4\alpha\xi]^{1/2}}{2\alpha^2\xi(\xi-1)[(\alpha-1)^2+4\alpha\xi]^{1/2}},
}[EqnrB]
\textit{for all} values of $\alpha=\epsilon^2/\epsilon_c^2$ and $\xi$, and they lead to
\eqna{
x(\alpha,\xi)&=\left[1+\frac{(\alpha-1)^2}{4\alpha\xi}\right]^{-1/2},\\
R(\alpha,\xi)&=1-\frac{1}{2(1-\xi)}\left[1+\frac{\alpha(1-2\xi)-1}{[(\alpha-1)^2+4\alpha\xi]^{1/2}}\right],
}[EqnxRsoln]
\textit{for all} values of $\alpha>0$ and $0<\xi\leq1$.  Here, $x=\bar{r}/\tilde{r}_\text{dS}$ is the ratio of the bubble radius $\bar{r}$ to the false dS vacuum radius $\tilde{r}_\text{dS}$ while $R$ is the ratio of the instanton action to the false dS vacuum background action defined before in \EqnR.  It is important to note that the functional dependence of the ratios $x$ and $R$ is completely fixed in the thin-wall approximation up to $\epsilon_c$, which depends only on the shape of the scalar potential under investigation \Eqnecsoln.  As explained before, due to the thermal nature of the bounce at large $\epsilon$, the form of $R$ is corrected when the inequalities,
\eqn{\alpha\ll\frac{1}{\epsilon_c^4},\quad\quad(\alpha-1)^2+4\alpha\xi\ll\frac{1}{\epsilon_c^4},}[EqnIneq]
are not satisfied.\footnote{The first inequality is modified to $\alpha\cdot\text{max}(\xi,1-\xi)\ll\frac{1}{\epsilon_c^4}$ when $\xi\notin(0,1]$, although \EqnxRsoln must be interpreted carefully in that regime.  As usual \EqnrB is valid for all spaces, not just dS spaces.}  Finally, it is interesting to note that the monstrous algebraic tangle mentioned in \cite{Coleman:1980aw} actually holds the key to the peculiar behavior of the bounce at large $\epsilon$, mainly the observed phase transition to which we now turn.

\subsubsection{Phase Transition}

It is interesting to point out first that the ratio of the bubble radius to the false dS vacuum radius in \EqnxRsoln is always smaller than or equal to $1$.  This is consistent since the bubble cannot be larger than the false dS vacuum into which it must fit.  In fact, $x$ reaches its maximum $x=1$ at $\alpha=1$ for all $\xi$ of interest.  Thus, at the critical value $\epsilon=\epsilon_c$, the bubble radius is as large as the radius of the false dS vacuum.  This observation offers another explanation why the instanton solution must decrease outside the bubble when $\epsilon>\epsilon_c$. Indeed, the bubble radius at the time of materialization in units of the false dS vacuum radius can only shrink from then on.\footnote{This statement is about $dx/d\alpha$ and has nothing to do with the growth of the bubble.  Once it materializes, the bubble always grows.}

\begin{figure}[!t]
\centering
\resizebox{16cm}{!}{
\includegraphics{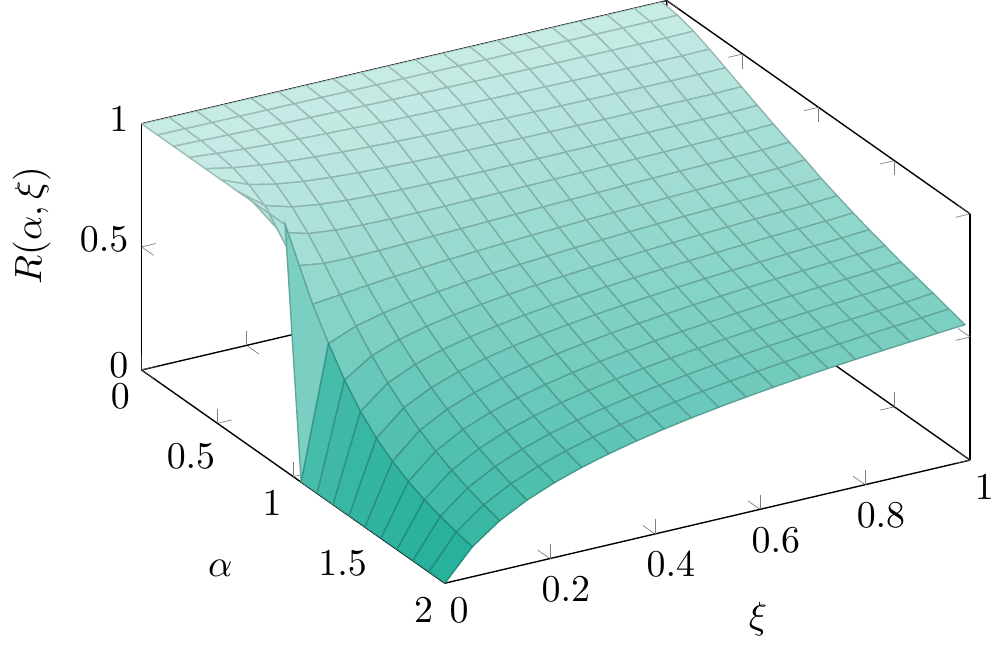}
\hspace{2cm}
\includegraphics{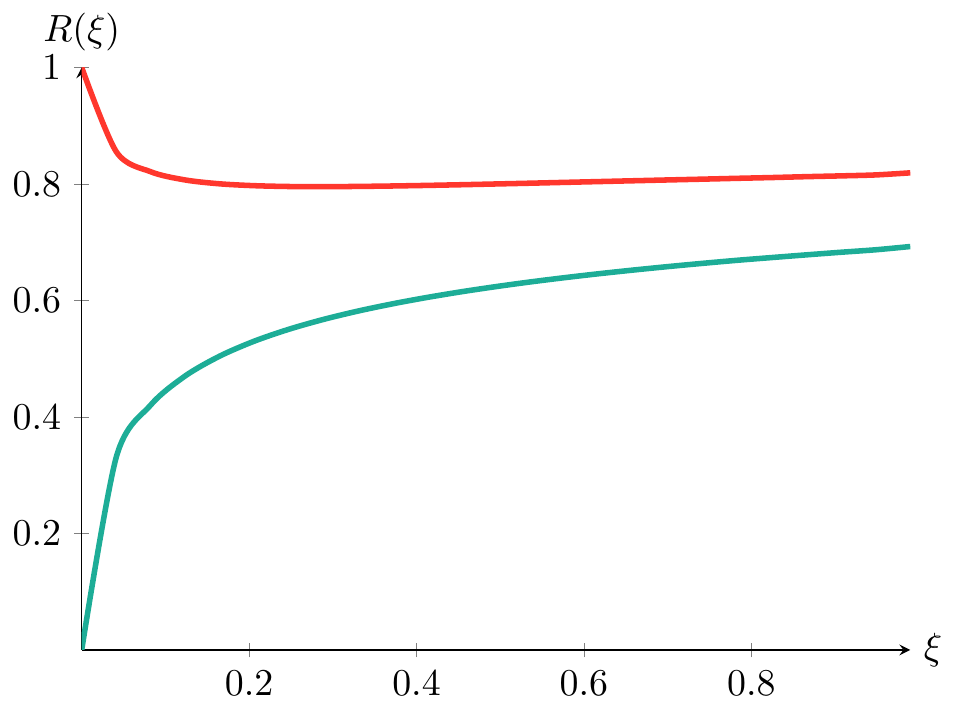}
}
\caption{$R$ as a function of $\alpha$ and $\xi$ (left) and $R$ as a function of $\xi$ for $\alpha=3/4$ (upper curve) and $\alpha=5/4$ (lower curve) (right).}
\label{FigR}
\end{figure}

Another important observation to make about \EqnxRsoln is that $R(\alpha,1)=\frac{2\alpha+1}{(1+\alpha)^2}$ for all $\alpha$.  Thus, for $\xi=1$ the ratio tends to $1$ when $\alpha\to0$ while it tends to $0$ when $\alpha\to\infty$.  The behavior of $R$ at $\xi=0$ is, however, strongly dependent on $\alpha$, being completely different when $\alpha$ is smaller or larger than $1$, as can be seen in figure \ref{FigR}.

When $0<\alpha<1$, $R$ behaves as
\eqn{R(\alpha<1,\xi)=\left\{\begin{array}{l}1-\frac{\alpha^2}{(1-\alpha)^2}\xi+\mathcal{O}(\xi^2)\\\frac{2\alpha+1}{(1+\alpha)^2}-\frac{\alpha^2(1-2\alpha)}{(1+\alpha)^4}(\xi-1)+\mathcal{O}[(\xi-1)^2]\end{array}\right.,}[EqnRas]
around $\xi=0$ and $\xi=1$, respectively.  Depending on the value of $\alpha$, there can be a minimum in the interval $\xi\in(0,1]$; nevertheless, $R$ is always larger than $1/2$.  More importantly, though, $R$ tends to $1$ when $\xi\to0$.  Therefore, the ratio $R$ is always large when $0<\alpha<1$.

When $\alpha=1$, we have
\eqn{R(\alpha=1,\xi)=1-\frac{1}{2(1+\sqrt{\xi})}.}[EqnRa]
In this case, $R$ is also always larger than $1/2$.  In fact, \EqnRa is consistent with what is observed for $\alpha<1$ when $\xi\to1$ as seen in \EqnRas.  However, since $R$ tends to $1/2$ when $\xi\to0$, \EqnRa behaves differently than \EqnRas, which tends to $1$ in this limit, suggesting a transition between two phases of $R(\alpha,\xi)$ as $\alpha$ increases.

Finally, the most interesting situation occurs when $\alpha>1$, where
\eqn{R(\alpha>1,\xi)=\left\{\begin{array}{l}\frac{2\alpha-1}{(\alpha-1)^2}\xi+\mathcal{O}(\xi^2)\\\frac{2\alpha+1}{(1+\alpha)^2}+\frac{\alpha^2(2\alpha-1)}{(1+\alpha)^4}(\xi-1)+\mathcal{O}[(\xi-1)^2]\end{array}\right..}[EqnRal]
Here, $R$ tends to $0$ as $\xi\to0$.  Indeed, although the behavior of $R$ for $\alpha>1$ is consistent with the behavior of \EqnRas and \EqnRa when $\xi\to1$, the ratio in \EqnRal is strongly suppressed as $\xi\to0$.  It is therefore possible to obtain very small ratios of $R$ when $\alpha>1$, a situation that was not possible for $0<\alpha\leq1$.  There is thus a transition for $R$ when $\alpha$ becomes larger than $1$.  Actually, since the behavior as $\xi\to1$ is the same for all $\alpha$ as already mentioned above, there is another transition between the regimes of large $\xi$ with large $R$ and the regime of small $\xi$ with small $R$ when $\alpha>1$.  Thus, for $\alpha>1$, there exists a ``critical point'', denoted by $\xi_c$, which divides the two regimes.  Although there are neither extrema nor inflection points in $R$ when $\alpha>1$, it is nevertheless possible to obtain a critical point $\xi_c$ by choosing the intersection point of the linear curves describing the two regimes \EqnRal, which leads to
\eqn{\xi_c=\frac{(\alpha-1)^2(6\alpha^2+4\alpha+1)}{(2\alpha-1)(6\alpha^3+5\alpha^2+4\alpha+1)}.}[Eqnxcsoln]
Although this definition is not unique (reflecting our ignorance on the full quantum gravity theory on dS space), it has appealing properties, discussed below.  This critical value tends to $0$ when $\alpha\to1$, as expected from \EqnRa, while it tends to $1/2$ when $\alpha\to\infty$.  Its behavior in those two regimes is given by
\eqn{\xi_c=\left\{\begin{array}{l}\frac{11}{16}(\alpha-1)^2+\mathcal{O}[(\alpha-1)^3]\\\frac{1}{2}-\frac{5}{6}\alpha^{-1}+\mathcal{O}(\alpha^{-2})\end{array}\right.,}[Eqnxc]
and $\xi_c$ is always smaller than $1/2$.  At the critical value, the ratio $R$ becomes
\eqn{R(\alpha>1,\xi_c)=\frac{20\alpha^3+5\alpha^2-6\alpha-3-(8\alpha^2+\alpha-1)\sqrt{\frac{(2\alpha-1)(3\alpha+1)(2\alpha^2+\alpha+1)}{(2\alpha+1)(6\alpha-1)}}}{4(\alpha+1)^2(3\alpha^2-1)},}[EqnRaxc]
which behaves as
\eqn{R(\alpha>1,\xi_c)=\left\{\begin{array}{l}\frac{15-2\sqrt{15}}{30}+\frac{97\sqrt{15}}{1800}(\alpha-1)+\mathcal{O}[(\alpha-1)^2]\\\alpha^{-1}+\mathcal{O}(\alpha^{-2})\end{array}\right.,}
around $\alpha=1$ and $\alpha=\infty$, respectively.  Hence, \EqnRaxc tends to a non-zero value as $\alpha\to1$ while it tends to $0$ as $\alpha\to\infty$.  Moreover, \EqnRaxc has a maximum for $\alpha>1$ and is always smaller than $1/2$, which was not possible for $0<\alpha\leq1$.

To recapitulate, there exists a phase transition in $R$ as $\xi\to0$ when $\alpha$ transitions between values smaller than $1$ to values larger than $1$.  In the former phase, $R$ is always large while in the latter phase $R$ can be made arbitrarily close to zero.  A possible way to determine the critical value $\xi_c$ below which $R$ in the latter phase is considered small is to choose $\xi_c$ as the point where the ratio $R$ transitions from large to small values.  These observations have important consequences for the tunneling rate per unit time per unit volume and the entropic interpretation of dS space.  This discussion thus ends the study of the ratio $R$ in the different regimes, although we remind the reader that corrections to $R$ must be taken into account when $\alpha\gg1$ and the inequalities \EqnIneq are not verified.  Moreover, contrary to the common lore, it is important to point out that the thin-wall approximation does display all the interesting behaviors of more generic potentials.

\subsubsection{Numerics}

Before deriving the constraints on the scalar potential, let us first extend the range of validity of \EqnxRsoln to more generic scalar potentials.  Indeed, although it is not possible to determine the ratio $R$ analytically for generic scalar potentials that are not of the thin-wall type, it is nevertheless possible to study it numerically.  As we show below, the simple analytical answer \EqnxRsoln for the ratio $R$ describes quite well such generic scalar potentials, although \Eqnecsoln does not lead to accurate $\epsilon_c$ in these cases.  This is not that surprising considering that the specifics of the scalar potential appear only in the determination of $\epsilon_c$ and nowhere else.  Moreover, although the three regions (inside the bubble, in the wall, and outside the bubble) are not as clearly defined as in the thin-wall approximation, the behavior of $R$ in terms of $\alpha$ and $\xi$ for generic potentials should be well parametrized by \EqnxRsoln as long as $\alpha$ and $\xi$ are not too large.

\begin{figure}[!t]
\centering
\resizebox{16cm}{!}{
\includegraphics{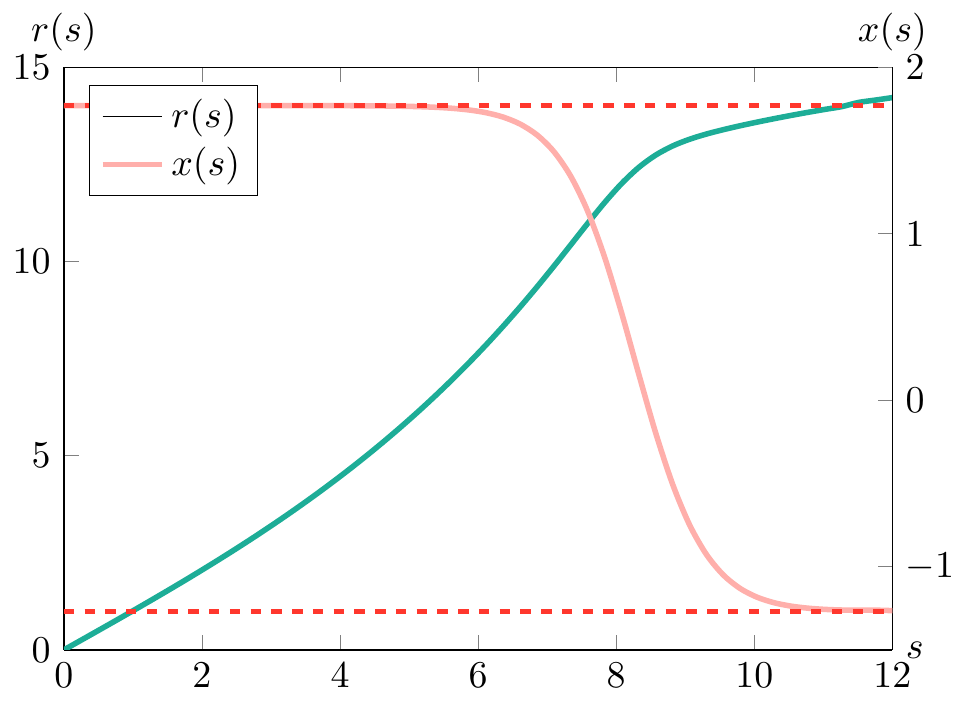}
\hspace{2cm}
\includegraphics{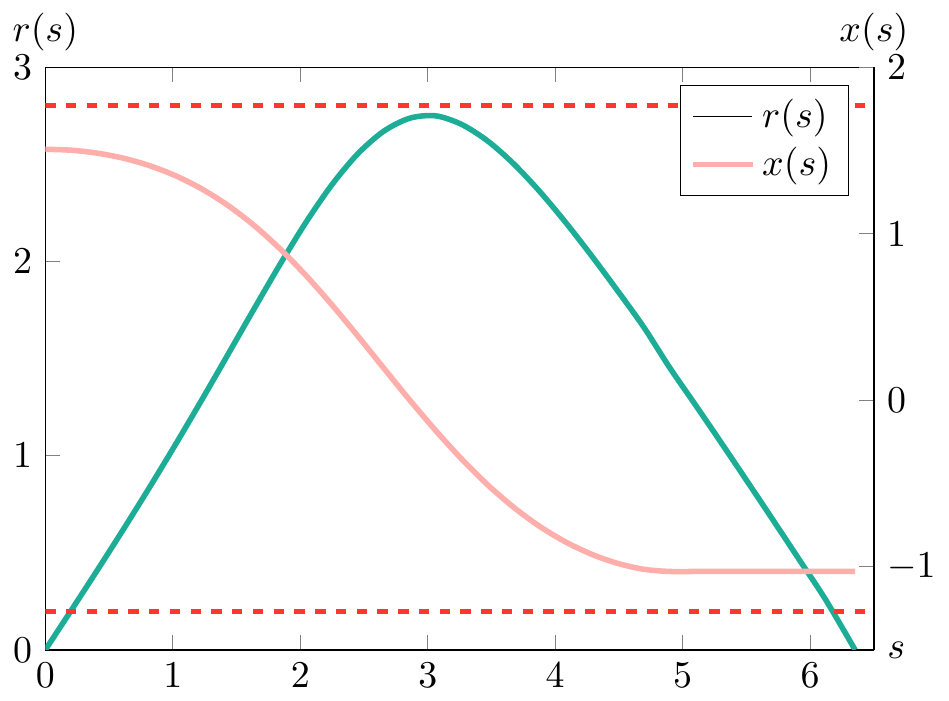}
}
\caption{Numerical solutions for $a=1.5$, $b=0.5$, and $\xi=0.1$.  The instanton solution in the left panel is obtained with $\epsilon=0.2$ and corresponds to a subcritical transition while the instanton solution in the right panel is obtained with $\epsilon=0.5$ and corresponds to a supercritical transition.}
\label{FigNumSol}
\end{figure}

\begin{figure}[!t]
\centering
\begin{subfigure}[b]{\textwidth}
\resizebox{16cm}{!}{
\includegraphics{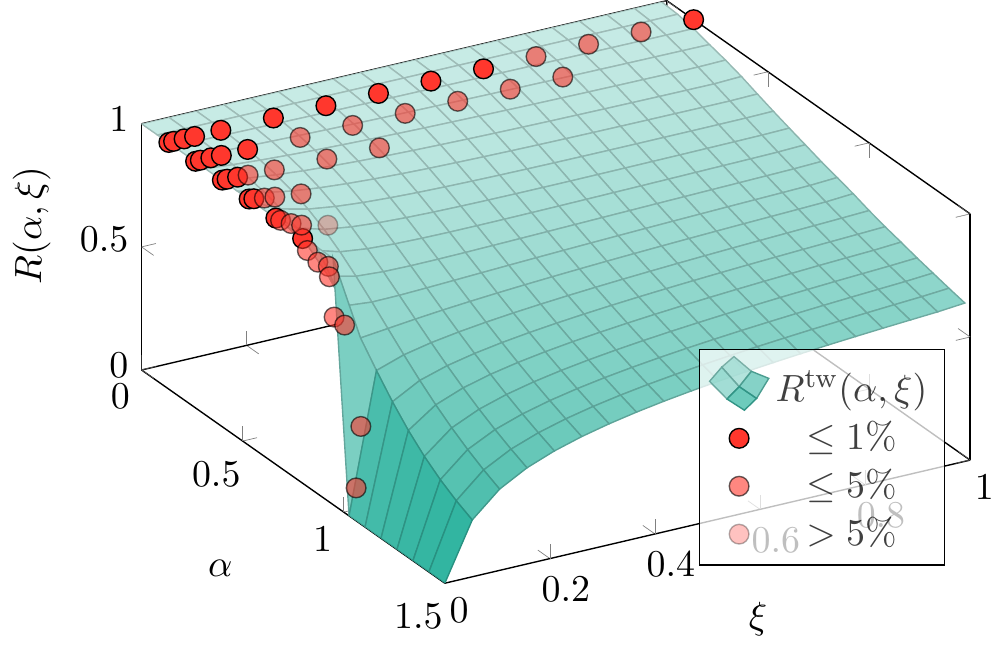}
\hspace{2cm}
\includegraphics{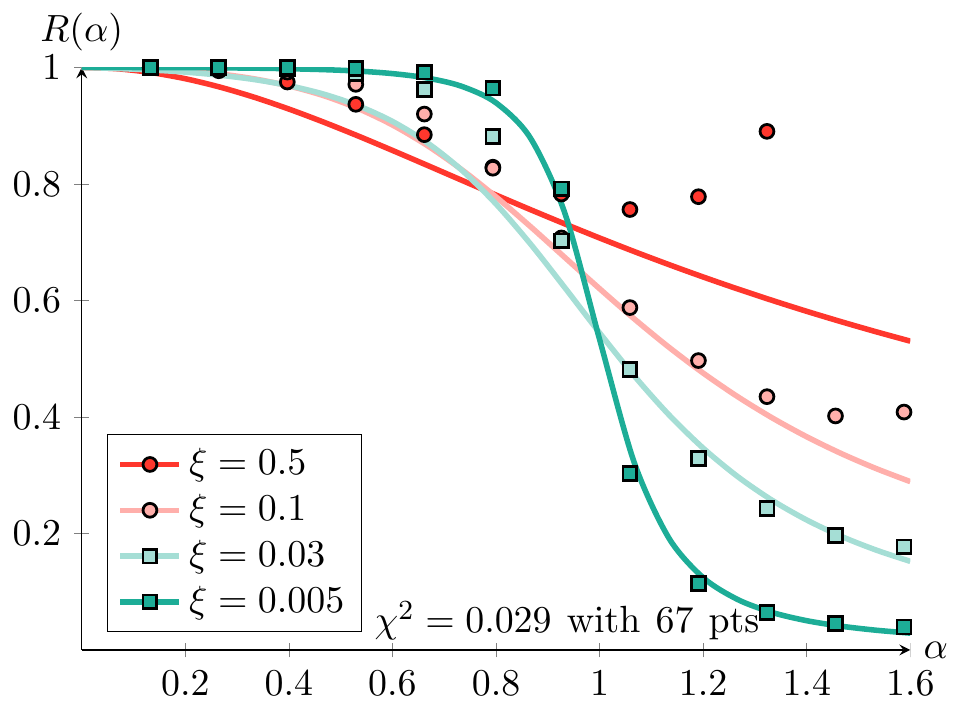}
}
\caption{Numerical ratios for $a=1.5$ and $b=1.0$ ($\epsilon_c^\text{fit}=0.37787\pm0.00001$).}
\end{subfigure}
\begin{subfigure}[b]{\textwidth}
\resizebox{16cm}{!}{
\includegraphics{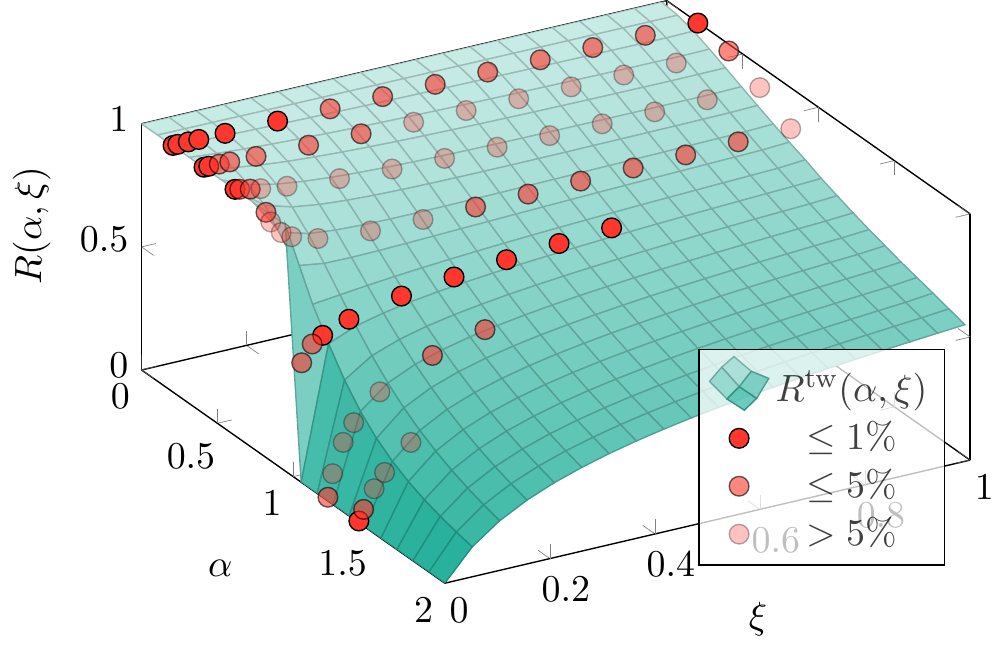}
\hspace{2cm}
\includegraphics{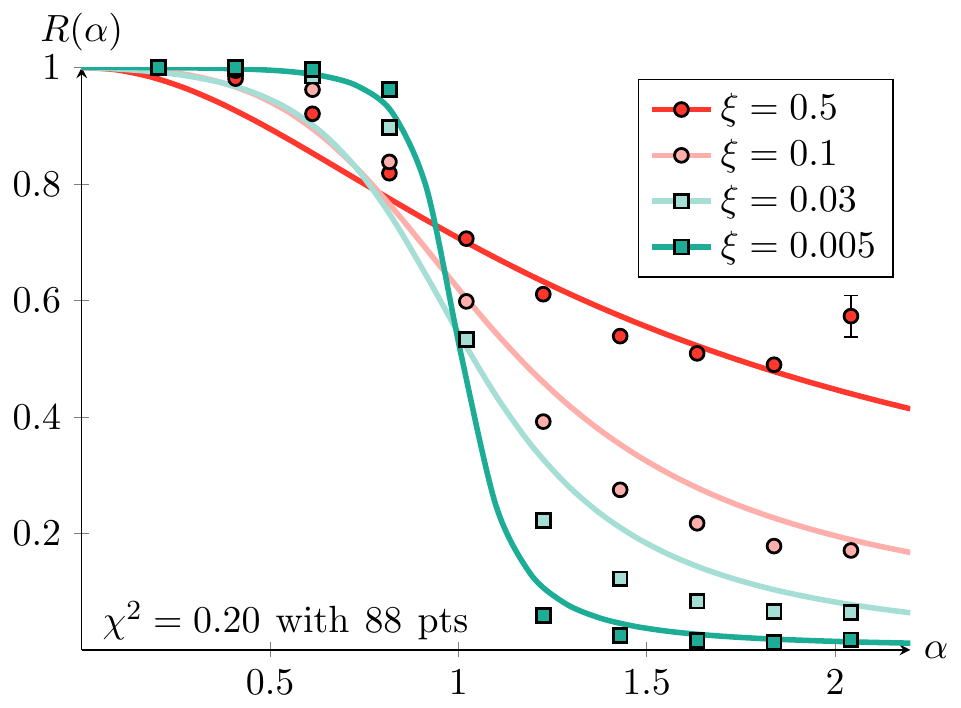}
}
\caption{Numerical ratios for $a=1.5$ and $b=0.5$ ($\epsilon_c^\text{fit}=0.2448\pm0.0001$).}
\end{subfigure}
\begin{subfigure}[b]{\textwidth}
\resizebox{16cm}{!}{
\includegraphics{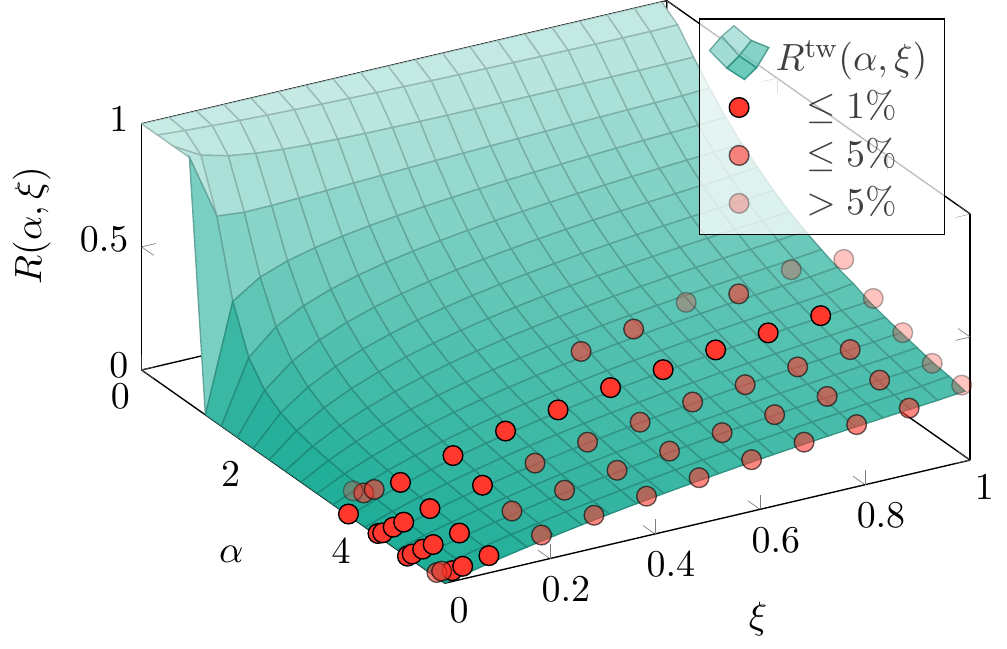}
\hspace{2cm}
\includegraphics{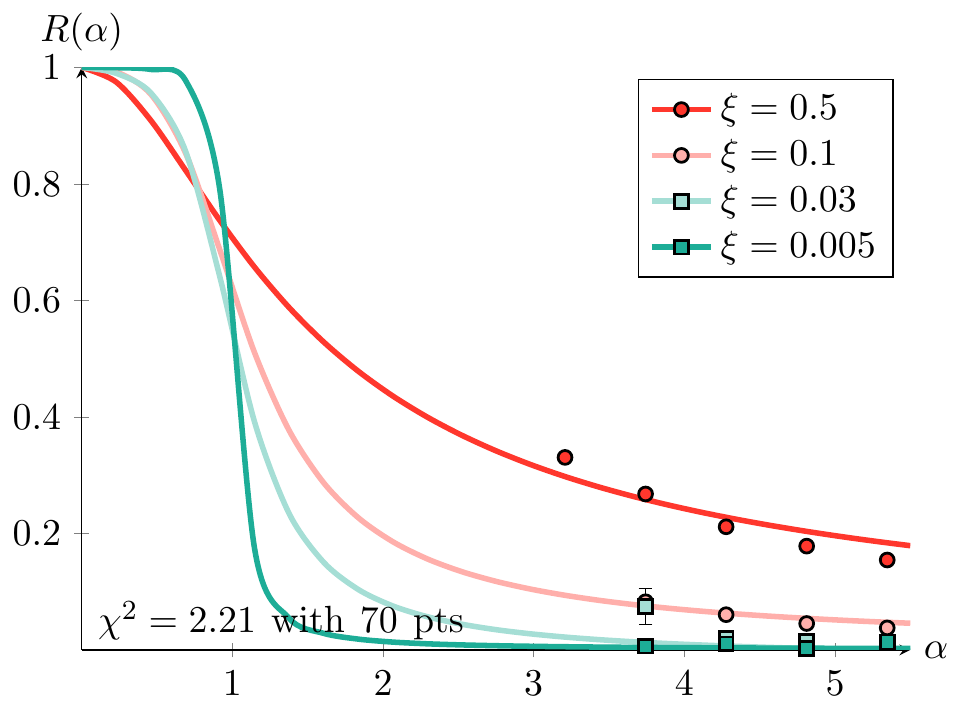}
}
\caption{Numerical ratios for $a=1.5$ and $b=0.1$ ($\epsilon_c^\text{fit}=0.054\pm0.002$).}
\label{FigRbTW}
\end{subfigure}
\caption{The left panels show the best fits of the thin-wall ratio $R^\text{tw}(\alpha,\xi)$ in green to the numerical solutions in red (the opacity gives the relative error to $R^\text{tw}$).  The missing points for large $\alpha$ are the thermal transitions, which are discarded for the fit.  The right panels show contour lines of $R^\text{tw}$ for different values of $\xi$.  The error bars are too small to appear (with typical relative error of $10^{-4}$).  Note the thermal effects for large $\alpha$ where the points tends to $1$ and the phase transition at $\alpha=1$.  Cases (a,b,c) differ by the value of the parameter $b$ and therefore $\epsilon_c^\text{fit}$, as indicated.}
\label{FigRbs}
\end{figure}

\begin{figure}[!t]
\centering
\resizebox{8cm}{!}{
\includegraphics{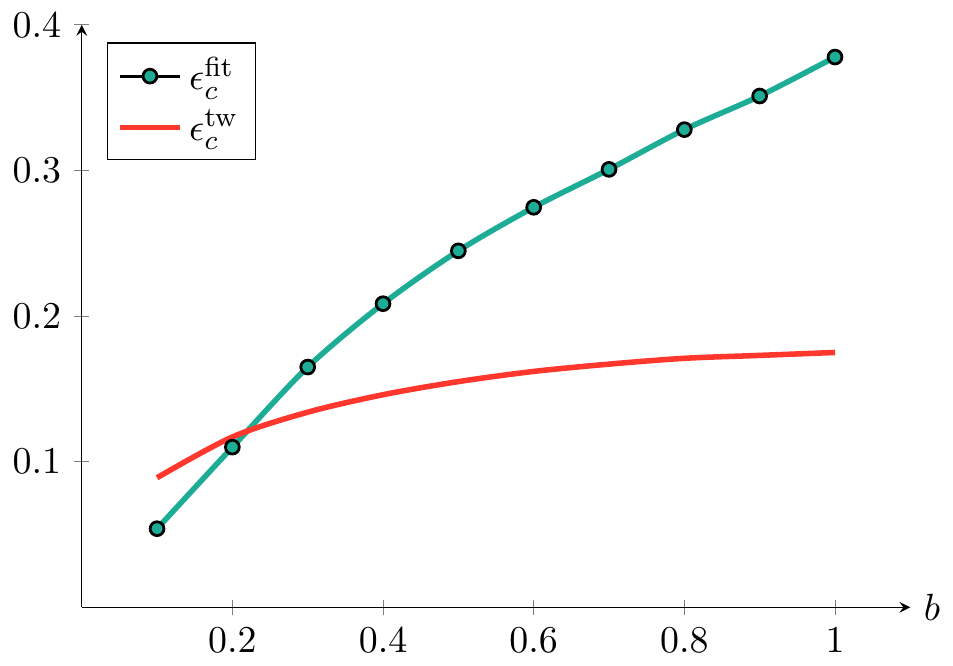}
}
\caption{Critical values $\epsilon_c^\text{fit}$ and $\epsilon_c^\text{tw}$ for $a=1.5$ and $b$ between $0.1$ and $1$.  For small $b$ the fitted $\epsilon_c^\text{fit}$ underestimates the correct critical value given by $\epsilon_c^\text{tw}$ since numerical instanton solutions are difficult to obtain numerically in the thin-wall approximation.  For larger $b$, $\epsilon_c^\text{tw}$ underestimates the correct critical value given by $\epsilon_c^\text{fit}$ since the thin-wall approximation breaks down.}
\label{FigEps}
\end{figure}

We investigated the behavior of $R$ numerically for three generic scalar potentials as functions of $\epsilon$ and $\xi$.  The three scalar potentials are given by \EqnPotential with $(a,b)=(1.5,1.0)$, $(a,b)=(1.5,0.5)$, and $(a,b)=(1.5,0.1)$ which correspond respectively to a scalar potential with a small energy barrier, a natural scalar potential with all variations of order one and a scalar potential with a small difference in the energy densities of the two vacua.  The results, along with fits to \EqnxRsoln, are shown in figures \ref{FigNumSol} and \ref{FigRbs} as well as a comparison between \Eqnecsoln and the fitted $\epsilon_c^\text{fit}$ in figure \ref{FigEps}.

In order to fit accurately the boundary conditions, we relied on the standard eight order Runge-Kutta method to integrate the equations of motion \EqnEOM combined with Brent's root finding algorithm, which seeks solutions whose scalar field is stable, \textit{i.e.} where $(dx/ds)|_{s_\text{max}}=0$.  Since $s_\text{max}$ is not fixed, a dynamical stop condition is introduced, so that integration is performed until the field is asymptotic to the false vacuum, its derivative becomes positive (so the instanton is not single-pass) or the bubble closes.  The sign reversal for the radius equation is dealt in the same way, which allows one to solve only for three first-order differential equations at once.  The algorithm thus minimizes the end slope of the scalar field by choosing the appropriate initial field value close to the true vacuum.  However, the downside of this approach is that it introduces fuzziness in the relation between the initial field value and the end slope value: the dynamical stop conditions amplify the numerical noise and introduce discontinuity in the relation, which prevents a precise computation of the initial conditions meeting the required criteria.  Indeed, the field is unstable near the false vacuum, so small perturbations make it diverge in one direction or the other until one of the stop criteria is met, therefore never staying asymptotic.  A workaround would be to also integrate the equations of motion backward from the false vacuum and fit the two solutions at a midpoint, but this is rendered computationally expensive by the dynamical time of integration.  Since we only require an estimate of the bounce, we resolve the initial field value up to the scale where the numerical errors take over (consistently about $10^{-7}$), and then average the bounce over a random sampling of the interval containing the root.  The asymptotic part is replaced by the dS solution.  Typical numerical solutions with $\epsilon$ under and over $\epsilon_c$ are shown in figure \ref{FigNumSol}.  Notice the qualitative difference between the two cases: when $\epsilon\leq\epsilon_c$, $r(s)$ and $dr/ds$ change weakly in the wall such that $dr/ds$ stays positive while when $\epsilon>\epsilon_c$, the weak variation in $r(s)$ in the wall is enough to induce a large variation in $dr/ds$ in the wall such that it reaches a maximum, thus forcing $dr/ds<0$ afterwards.  Obviously, $dr/ds$ also changes sign when $\epsilon\leq\epsilon_c$ but only outside the wall.

We then evaluate $\epsilon_c$ by fitting $R$ for each point to the thin-wall curve \EqnxRsoln, which works surprisingly well even for thick-wall potentials (except for transitions with large $\alpha$ and/or $\xi$, as expected).  The uncertainty on $\epsilon_c$ is evaluated by a bootstrapping method: we generate many sets of points $R+\Delta R$, where $\Delta R$ is distributed according to variance found previously for each point, and we compute the variance of $\epsilon_c$ for these sets.  This yields the statistical error of the method, but does not account for the systematic error due to the distribution of valid solutions.  In particular, our algorithm underestimates slightly $\epsilon_c$ for the thin-wall potentials, since the solutions for low $\epsilon$ in the thin-wall limit stay close to the true vacuum for a very long time before decaying, therefore requiring great numerical precision.  This prevents the program from finding these solutions, which gives higher weight to the supercritical solutions.  Fortunately, this is the region where the analytical methods work best, so we can compare the two methods and estimate the numerical bias.

The fitted curves $R^\text{tw}$ are shown in figure \ref{FigRbs} for the three generic cases.
Note that, surprisingly, the thin-wall curve \EqnxRsoln fits the numerical solutions accurately even when $\Delta v$ is order one, except for large values of $\alpha$.  In this region, the instanton is thermal (see the right panel of figure \ref{FigNumSol}), so the instanton interpolates between excited states.  Such a transition occurs when the bubble closes too fast for a complete transition between both minima to occur, that is, when the gravitational effects are large.  The exact prescription for ignoring these solutions when searching $\epsilon_c^\text{fit}$ is ill-defined since the solutions always start in the vicinity of the true vacuum, but we found that requiring that $|x_0-x_T|/|x_F-x_T|\lesssim10^{-2}$ gave satisfying results in most cases.  The points used for the fit with their relative errors are shown on the left, while all the points for a given $\xi$ are shown on the right.  Notice also how the curve in figure \ref{FigRbTW} seems to underestimate $\epsilon_c$.  Indeed, the thin-wall estimate yields $\epsilon_c^\text{tw}\approx0.089$, which is slightly larger than the numerical value $\epsilon_c^\text{fit}=0.054\pm0.002$.  To allow a comparison, the values of $\epsilon_c^\text{fit}$ and $\epsilon_c^\text{tw}$ are shown for different $b$ in figure \ref{FigEps}.
Here $b$ is a measure of the vacuum energy difference between the two vacua compared to the height of the potential.  From figure \ref{FigEps} it is clear that the fitted value of $\epsilon_c^\text{fit}$ underestimates the correct value obtained from the thin-wall approximation until the ratio of the vacuum energy difference and the height of the potential reaches around $0.4$ while the thin-wall result $\epsilon_c^\text{tw}$ underestimates the correct value obtained from the fit above this ratio where the thin-wall approximation breaks down.  Since the fit looks even better for these latter cases, we are confident that the numerical results are accurate if somewhat slightly biased.  It is also important to point out that we obtained numerically $\epsilon_c^\text{fit}$ without relying on the fit to \EqnxRsoln, which is only possible outside the thin-wall regime due to numerics, and that these values match the ones shown in figure \ref{FigEps}.

Finally, if the scalar potential can be properly approximated in the region of interest for the instanton solution, it is possible to compute numerically the value of $\epsilon_c$ for a generic potential where the true minimum is far away in field space and does not play a role in finding the instanton solution.  For example, if the potential can be approximated by
\eqn{v(x)=\frac{a'x^n}{n}-\frac{b'x^p}{p},}[Eqnvapprox]
where $p>n\in\mathds{N}$, $a'$ and $b'$ are larger than $0$, the false vacuum is at $x_F=0$ and the true vacuum is at large $x$, then the following change of variables
\eqn{s=\frac{c'^{n/2-1}}{\sqrt{a'}}s',\quad\quad r=\frac{c'^{n/2-1}}{\sqrt{a'}}r',\quad\quad x=\frac{x'}{c'},\quad\quad\epsilon=c'\epsilon',}[EqnChange]
with $c'=(nb'/pa')^{1/(p-n)}$, leads to the same equations of motion \EqnEOM in terms of the primed quantities and
\eqn{v'(x')=\frac{x'^n}{n}(1-x'^{p-n}).}[Eqnvapproxchange]
Then one obtains numerically the instanton solutions for \Eqnvapproxchange given in table \ref{TabEps}.
\begin{table}[!t]
\centering
\resizebox{5cm}{!}{
\begin{tabular}{|cc|cc|}
\hline
$p$ & $n$ & $\epsilon_c'$ & $x_0'$\\
\hline
$2$ & $1$ & $0.60980$ & $4.36697$\\
$3$ & $1$ & $0.70688$ & $4.71565$\\
$3$ & $2$ & $0.84063$ & $5.29815$\\
$4$ & $1$ & $0.79998$ & $5.04952$\\
$4$ & $2$ & $0.94278$ & $5.49919$\\
$4$ & $3$ & $1.072$ & $4.96271$\\
\hline
\end{tabular}
}
\caption{Critical values $\epsilon_c'$ and initial points $x_0'$ for generic scalar potentials with true minima far away in field space that play no role in obtaining instanton solutions.}
\label{TabEps}
\end{table}
Thus, as long as the fit is satisfactory in the region between the false vacuum at $x=x_F=0$ and the beginning of the instanton at $x=x_0$, the critical value $\epsilon_c$ is easily obtained (as well as $x_0$) from table \ref{TabEps}, \EqnChange, and the appropriate fit to \Eqnvapprox.  

\subsection{Constraints on the Scalar Potential and Consistency Conditions}\label{SsConstraints}

The two constraints on the scalar potential come from the observation that the behavior of $R$ changes drastically when $\epsilon$ is smaller or larger than $\epsilon_c$, given by \Eqnecsoln in the thin-wall approximation.  As already described, for $\epsilon\leq\epsilon_c$ the ratio $R\approx\mathcal{O}(1)$ for all values of $\xi$ (more precisely $R\geq1/2$) while for $\epsilon>\epsilon_c$ the ratio $R$ tends to zero when $\xi\to0$.  This specific behavior was observed numerically before.  Indeed, it was shown in \cite{Banks:2005ru,Aguirre:2006ap,Bousso:2006am} that there exist scalar potentials for which the Euclidean action for the instanton stays finite as the false vacuum energy density is brought to zero from above.  In \cite{Aguirre:2006ap} it was argued that a consistent theory of quantum dS space must satisfy this condition, $\epsilon>\epsilon_c$, which is our first constraint on a generic scalar potential.  Indeed, when this particular situation occurs, the background Euclidean action blows up as $\xi\to0$ and it can be argued that the CDL instanton is a sub-Poincar\'e recurrence for small enough ratio $R$ \cite{Aguirre:2006ap}.  The constraint was used in \cite{Banks:2009nz} to argue against metastable SUSY breaking minima but was never used in parallel with RG flow techniques to constrain fundamental scalar masses. 

Our second constraint, which as far as we know is new, restricts $\xi$ such that the size of the ratio $R$ is small when the first constraint is satisfied (\textit{i.e.} $R<1/2$).  To obtain such a constraint, the critical value is chosen as the meeting point of the curves describing the behavior of $R$ around $\xi=0$ and $\xi=1$ when $\epsilon>\epsilon_c$, \textit{i.e.} $\xi<\xi_c$, for all scalar potentials, although the critical value is obtained from $R$ computed in the thin-wall approximation.  This definition is sensible since \EqnxRsoln describes the ratio $R$ well for generic potentials as long as $\alpha$ is not too large.  For large $\alpha$, our numerical analysis suggests $\xi\ll\xi_c$ instead and thus makes $\xi_c$ a conservative bound.  In any case, the answer depends on $\epsilon_c$ and is given by \Eqnxcsoln.  This definition, although somewhat arbitrary due to the lack of spectral features in $R$, makes sense since the ratio $R$ for $\epsilon>\epsilon_c$ and $\xi<\xi_c$ is always smaller than any $R$ with $\epsilon\leq\epsilon_c$.  Moreover, $\xi_c$ increases when the height of the scalar potential barrier increases, as expected physically.

Consequently, we obtain $B^\text{c}\approx\mathcal{S}_\text{dS}$, \textit{i.e.} $R$ small, when both
\eqn{\epsilon>\epsilon_c\hspace{1cm}\text{and}\hspace{1cm}\xi<\xi_c,}[EqnConstraints]
for generic scalar potentials.  In terms of the parameters of the potential, \EqnConstraints roughly becomes
\eqn{|\Delta\phi|\sqrt{1+c_\epsilon\frac{V_H-V_F}{\Delta V}}\gtrsim m_P\hspace{1cm}\text{and}\hspace{1cm}\rho_\text{vac}\lesssim\frac{c_\xi}{2}\Delta V,}[EqnConstraintsPotential]
where $V_H$ is the value of the scalar potential at the top of the barrier $\phi_H$, $\rho_\text{vac}=V_F$ and $c_\epsilon$ and $c_\xi$ are $\mathcal{O}(1)$ coefficients.  Thus, for a consistent theory of quantum dS space to exist, the finiteness of the Hilbert space suggests the constraints \EqnConstraints or \EqnConstraintsPotential on the scalar potential, which translate approximatively into the following: the distance in field space between the de Sitter vacuum and any other vacuum with negative cosmological constant must be of the order of the reduced Planck mass or larger; and the fourth root of the vacuum energy density of the de Sitter vacuum must be smaller than the fourth root of the typical scale of the scalar potential.\footnote{It is interesting to note that an entropic interpretation of the decay is also consistent for non-generic potentials where $V_H$ is very large in Planck units.}  Although the constraints have no direct relation to the large additive quantum corrections to fundamental scalar masses and the cosmological constant that lead to the hierarchy problems, it is always possible to posit them as consistency conditions.

Indeed, in the usual framework for the naturalness of small fundamental scalar masses and a small cosmological constant, where cut-off regularization is used, the large additive quantum corrections, of the order of the Planck scale, are obtained by extrapolating the behavior of QFT to very high energies where gravity becomes important.  Since quantum gravity in dS space is poorly understood, it is not too far-fetched to assume that an as-yet-unknown mechanism at the Planck scale dictates the large additive quantum corrections are red herrings that should be dismissed.  Therefore, it is plausible that a natural scalar potential instead looks like one that actually satisfies the constraints \EqnConstraints or \EqnConstraintsPotential.  Note, however, that the first constraint has nothing specific to say about the masses of fundamental scalars: an analysis of the scalar potential from the RG flow is needed to obtain a concrete consistency condition.  The second constraint, on the other hand, directly states upper bounds on the cosmological constant, the number of such bounds depending on the number of accessible vacua, shedding a new light on the corresponding hierarchy problem.  Our approach, demanding a consistent understanding of tunneling rates for dS spaces, leads to concrete constraints on scalar potentials that are necessary for an entropic understanding of quantum dS space, but arguing that such scalar potentials are consistent is merely a speculation.  Finally, since generic scalar potentials are consistent if they satisfy \EqnConstraints or \EqnConstraintsPotential, there is no such thing as a fine-tuning measure in our framework.

The previous analysis can be restated directly in terms of the relevant timescales.  Indeed, the timescale $t_\text{CDL}$ for the CDL ``tunneling'' event to occur can be straightforwardly expressed in terms of the Poincar\'e recurrence time $t_\text{PR}$ (up to a polynomial prefactor related to the determinant) as
\eqn{\frac{t_\text{CDL}}{R_\text{dS}}\approx\left(\frac{t_\text{PR}}{R_\text{dS}}\right)^{1-R}.}[EqnTdS]
It is clear from \EqnTdS that $R$ must be parametrically small ($R\ll1$) for that timescale to be of the order of (but always smaller than) the Poincar\'e recurrence time, enabling an entropic interpretation of the ``decay''.  The constraints \EqnConstraints lead to a conservative bound on the ratio $R$ given by $R<1/2$.  Thus, with that bound the CDL timescale is still exponentially smaller than the Poincar\'e recurrence time, explaining why it is only a conservative bound.  Nevertheless, with that conservative bound the CDL timescale is still exponentially larger than the typical timescale $t_\text{BH}$ for the nucleation of a dS black hole of maximal size, which is approximatively given by \cite{Ginsparg:1982rs}
\eqn{\frac{t_\text{BH}}{R_\text{dS}}\approx\left(\frac{t_\text{PR}}{R_\text{dS}}\right)^{1/3}.}[EqnTBH]
From \EqnTBH, a dS black hole of maximal size nucleates in a quantum dS space with a scalar potential satisfying \EqnConstraints an exponential number of times before a CDL instanton occurs.  Hence, the conservative bound on the ratio $R$ deduced from \EqnConstraints, more specifically from the bound on $\xi$, which was somewhat arbitrary due to a lack of spectral features in the ratio, leads to a timescale exponentially longer than the typical timescale for black hole nucleation, strongly suggesting that CDL instantons are not problematic when the constraints \EqnConstraints are satisfied.

\subsubsection{Comparison with Hawking-Moss Instantons}

It is interesting to consider whether Hawking-Moss instantons modify the previous discussions, which were based on CDL instantons instead.  For this purpose, it is enlightening to first compare \EqnxRsoln to the corresponding result for the HM instanton \cite{Hawking:1981fz}, for which the relevant ratio is
\eqn{R^\text{HM}(\eta,\xi)=\frac{S_E(\phi_H)}{S_E(\phi_F)}=\frac{\xi}{\xi+\eta},}[EqnRHM]
where $\phi_H=Mx_H$ is the location of the top of the barrier and $\eta$ is the height of the scalar potential barrier in units of the energy difference between the two vacua, \textit{i.e.} $v_H\equiv v(x_H)=(\xi+\eta)\Delta v$.  For any $\eta$, the ratio \EqnRHM behaves as
\eqn{R^\text{HM}(\eta,\xi)=\left\{\begin{array}{l}\frac{1}{\eta}\xi+\mathcal{O}(\xi^2)\\\frac{1}{1+\eta}+\frac{\eta}{(1+\eta)^2}(\xi-1)+\mathcal{O}[(\xi-1)^2]\end{array}\right.,}[EqnRHMxi]
around $\xi=0$ and $\xi=1$, respectively.  Thus, $R^\text{HM}$ tends to $0$ as $\xi\to0$.  Therefore, the HM bounce is sub-dominant compared to the CDL bounce when $\epsilon\leq\epsilon_c$, strengthening the necessity of the first consistency condition mentioned above.  When $\epsilon>\epsilon_c$, the HM bounce is sub-dominant compared to the CDL bounce when
\eqn{\eta\geq\frac{(\alpha-1)^2}{2\alpha-1},}
and dominant compared to the CDL bounce when
\eqn{\eta<\frac{(\alpha-1)^2}{2\alpha-1}.}
Thus, for a very small height of the scalar potential barrier, it is possible that the dominant quantum tunneling channel is through HM instantons.  However, an entropic interpretation of the decay is always possible since the ratio $R^\text{HM}$ can be made arbitrarily small as $\xi\to0$, \textit{i.e.} as the false vacuum cosmological constant is taken to zero.  Since the ratio $R^\text{HM}$ does not exhibit extrema or inflection points in the region $\xi\in[0,1]$, to obtain the critical value for $\xi$, it is again possible to choose the intersection point of the linear curves \EqnRHMxi describing the ratio \EqnRHM in the two regimes around $\xi=0$ and $\xi=1$, respectively, which leads to
\eqn{\xi_c^\text{HM}=\frac{\eta}{1+2\eta}.}[EqnxHMc]
This critical value tends to $0$ when $\eta\to0$ and to $1/2$ when $\eta\to\infty$.  At the critical value \EqnxHMc, the ratio $R^\text{HM}$ is
\eqn{R^\text{HM}(\eta,\xi_c^\text{HM})=\frac{1}{2(1+\eta)},}[EqnRHMaxc]
which tends to $1/2$ when $\eta\to0$ and $0$ when $\eta\to\infty$.  Moreover, \EqnRHMaxc is always smaller than $1/2$.  Therefore, the critical value $\xi_c^\text{HM}$ is smaller than $1/2$, for which the ratio $R^\text{HM}$ is also smaller than $1/2$, in accord with the second consistency condition already mentioned.

Hence, the general conclusions implied by \EqnConstraints prevail when the (usually sub-dominant) contributions from HM quantum tunneling events are taken into account.


\section{Quantum dS Space and the SM}\label{SSM}

It is interesting to study the constraints \EqnConstraints on the SM and its extensions to determine what can be learned from this train of thought.\footnote{An alternative take on the interplay between quantum gravity in dS space and the metastability of the SM electroweak vacuum is considered in \cite{Espinosa:2015qea}.}  It is actually possible to do high-precision numerics for \EqnConstraints using the SM effective action and obtain concrete bounds on the SM Higgs mass and the SM vacuum energy density.  Indeed, the same methods developed above can be applied to the SM Higgs potential to obtain a lower bound on the Higgs mass.  Specifically, when the critical SM Higgs potential occurs close to the stability region, the thin-wall approximation is used.  When the true minimum lies far away in field space, the results of table \ref{TabEps} are more convenient.  When such particular situations do not occur, a full numerical analysis is performed to determine at which Higgs mass the SM Higgs potential becomes overcritical.
\begin{figure}[!t]
\centering
\resizebox{8cm}{!}{
\includegraphics{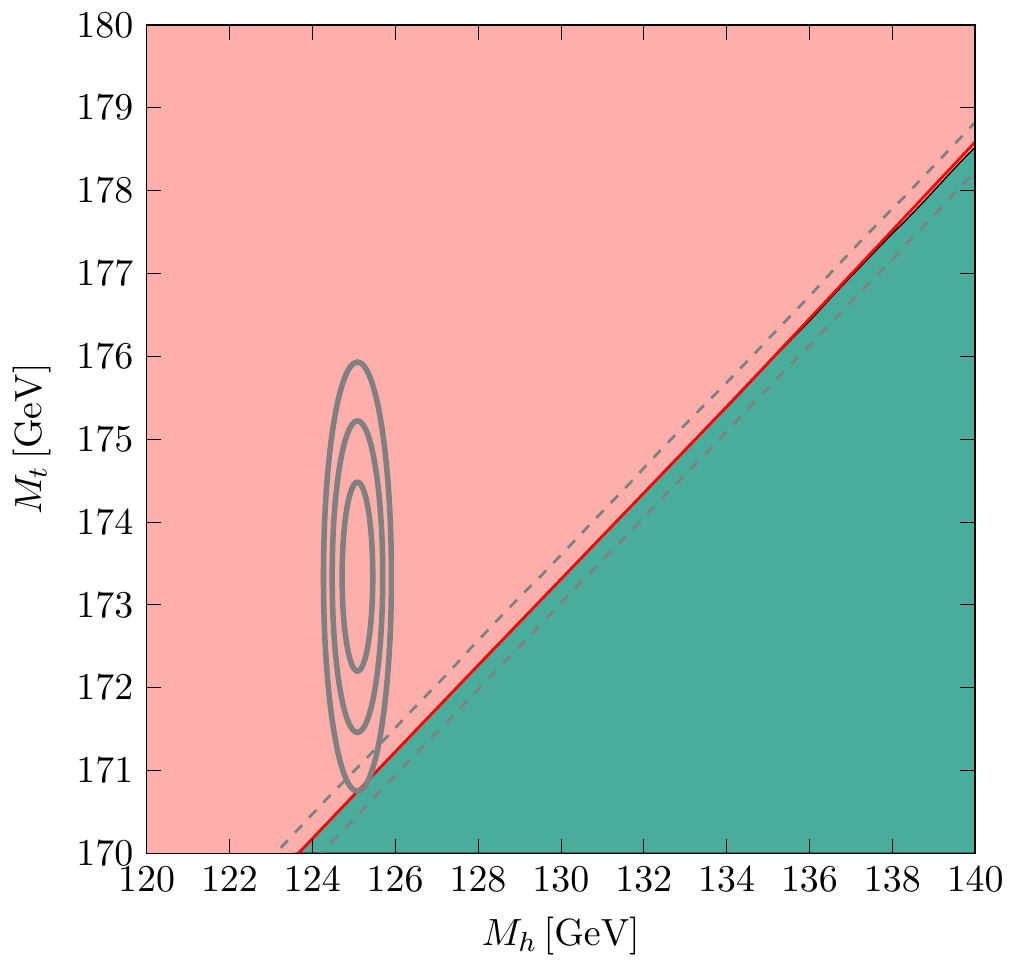}
}
\caption{In the red area the SM EW vacuum is only metastable while in the green area it is stable (at least up to the Planck scale).  The $1\sigma$ ($68\%$ C.L.) to $3\sigma$ ($98\%$ C.L.) ellipses correspond to the measured values of the top ($M_t=173.34\pm0.76\,\text{GeV}$ \cite{ATLAS:2014wva}) and Higgs ($M_h=125.09\pm0.24\,\text{GeV}$ \cite{Aad:2015zhl}) masses.  The dashed lines show the error in the stability boundary due to $\alpha_s(M_Z)=0.1184\pm0.0007$ \cite{Bethke:2012jm}.}
\label{FigSMStab}
\end{figure}

Thus, the discussion of the previous sections turns out to be relevant in the context of the SM.  In fact, for Higgs boson and top mass values near the measured ones (and assuming no new physics up to $M_P$), the electroweak (EW) vacuum is not the global minimum of the SM effective potential, which develops an instability at very high field values \cite{EliasMiro:2011aa,Bezrukov:2012sa,Degrassi:2012ry,Buttazzo:2013uya}, with an AdS minimum appearing at even higher fields.  The potential has this metastable EW vacuum in the red region of the $(M_h,M_t)$ plane shown in figure \ref{FigSMStab}, while it is stable in the green region.  The figure also indicates the values of $M_h$ and $M_t$ measured by experiment [with $1\sigma$ ($68\%$ C.L.) to $3\sigma$ ($98\%$ C.L.) error ellipses].  In view of this, it is of clear interest to ask if such SM potentials are compatible with the dS quantum gravity constraints (and if so, for what range of masses).  In other words, where does the curve $\epsilon=\epsilon_c$ lie in the $(M_h,M_t)$ plane of figure \ref{FigSMStab} ?

The SM potential at large field values is well approximated as\footnote{The quartic coupling in this formula differs from the usual running quartic coupling by finite radiative corrections that are nevertheless small \cite{EliasMiro:2011aa,Bezrukov:2012sa,Degrassi:2012ry,Buttazzo:2013uya}.} $V(h)\simeq\lambda(\mu=h)h^4/4$ \cite{EliasMiro:2011aa,Bezrukov:2012sa,Degrassi:2012ry,Buttazzo:2013uya}.  Its shape is dictated by the running of the Higgs quartic coupling $\lambda(\mu)$, which one should evaluate at a renormalization scale $\mu\sim h$, as indicated.  In this language, the potential instability mentioned before results from $\lambda(\mu)$ becoming negative at some high scale $\mu=\Lambda_I$, as a result of sizeable top loop corrections (see \cite{EliasMiro:2011aa,Bezrukov:2012sa,Degrassi:2012ry,Buttazzo:2013uya} for a detailed state-of-the-art next-to-next-to-leading-order discussion).

Inside the metastability region (the red area in figure \ref{FigSMStab}) the quartic coupling becomes negative at some high scale $\Lambda_I$ well below the Planck scale (\textit{e.g.}~at around $\Lambda_I\sim10^{10}\,\text{GeV}$ for the central values of $M_h$ and $M_t$).  For $h>\Lambda_I$, the SM potential becomes lower than the EW vacuum and keeps getting lower and lower for higher field values until eventually $\lambda$ starts to grow towards positive values again (when positive corrections from gauge boson loops overcome the top loop effect, that becomes weaker in the UV as the top Yukawa gets smaller and smaller), creating in the potential an additional AdS minimum.  This minimum normally appears for $h>M_P$.  Instead, in the green
area, $\lambda(\mu)$ stays positive all the way up to $M_P$ leading to a stable potential.
\begin{figure}[!t]
\centering
\resizebox{12cm}{!}{
\includegraphics{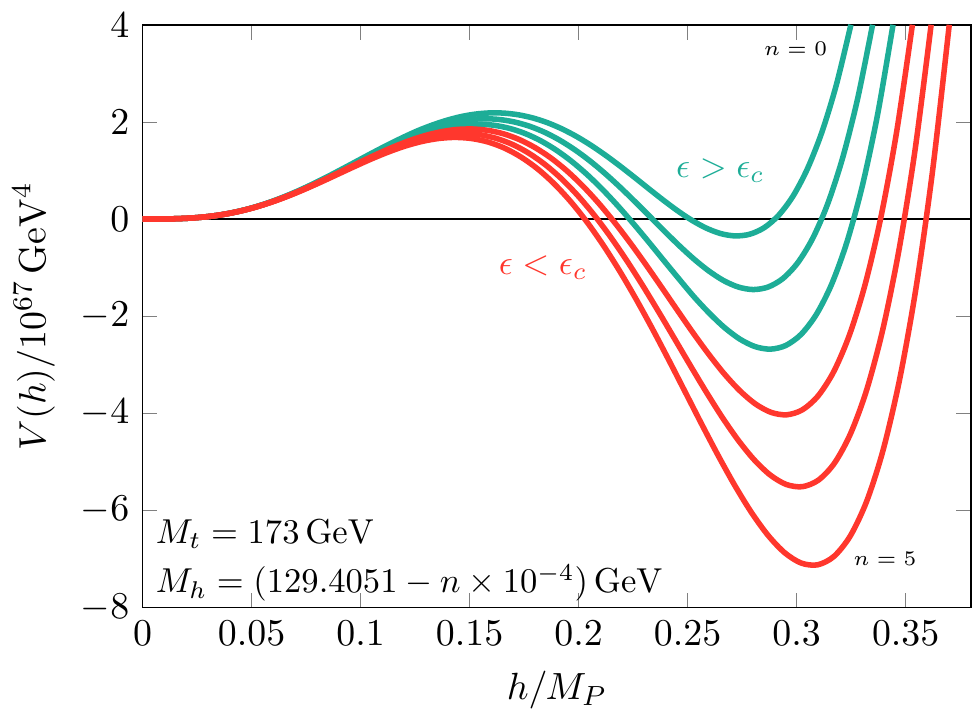}
}
\caption{SM effective potential for $M_t=173\,\text{GeV}$, $\alpha_s(M_Z)=0.1184$, and Higgs mass values very close to the stability bound (from below), as indicated.  Potentials that have $\epsilon>\epsilon_c$ ($\epsilon<\epsilon_c$) are in green (red) with $\epsilon/\epsilon_c=\{1.63,1.25,1.05,0.91,0.82,0.75\}$ from $n=0$ (for $M_h=129.4051\,\text{GeV}$) to $n=5$ (for $M_h=129.4046\,\text{GeV}$).}
\label{FigSMGD173}
\end{figure}

Concerning the location of $\epsilon=\epsilon_c$, typical potentials well inside the metastability region have very deep AdS minima with very shallow barriers and thus have $\epsilon<\epsilon_c$.  In order to get potentials with $\epsilon>\epsilon_c$ one needs to get close to the stability boundary that separates metastable and stable potentials.  In fact, the shape of the potentials in the neighbourhood of the stability line is extremely sensitive to $M_h$ and $M_t$.  To understand why, note that the stability boundary corresponds to a very special running of $\lambda(\mu)$, that touches zero at some high scale $\mu=\Lambda_I$ and goes up to positive values again.  That is, it corresponds to $\lambda=0$, $d\lambda/d\mu=0$ at $\mu=\Lambda_I$.  The potential corresponding to this boundary value has two degenerate minima,\footnote{Here, we are neglecting the small cosmological constant at the EW vacuum.  Strictly speaking, degeneracy of both minima requires $\lambda(\Lambda_I)$ to be extremely small but non-zero.} the EW one and the one at $h\simeq\Lambda_I$.

A tiny change in $M_h$ or $M_t$ away from the boundary values either makes $\lambda$ near $\Lambda_I$ positive or negative, modifying dramatically the potential shape. For $\lambda(\mu\sim\Lambda_I)>0$ the minimum near $\Lambda_I$ quickly goes up and disappears while for $\lambda(\mu\sim\Lambda_I)<0$ it instead becomes very deep.  This behaviour is illustrated, for $M_t=173\,\text{GeV}$ and $\alpha_s(M_Z)=0.1184$, in figure \ref{FigSMGD173}, which shows the SM potential for a series of $M_h$ values very close to the stability boundary.  Note that
the step between the different mass values chosen is very small, $\Delta M_h=10^{-4}\,\text{GeV}$ (smaller than the Higgs total width $\Gamma_h\sim4\times10^{-3}\,\text{GeV}$).  The value of $\epsilon/\epsilon_c$ for these potentials grows when $M_h$ increases towards stability, eventually becoming larger than $1$.  We have plotted in green (red) those potentials that have $\epsilon>\epsilon_c$ ($\epsilon<\epsilon_c$), with the precise values given in the figure caption.

We therefore see that the specific location where $\epsilon=\epsilon_c$ is very close to the stability boundary.  This can be appreciated in figure \ref{FigSMStab}, which shows the $\epsilon=\epsilon_c$ line in red, nearly on top of the stability line.  The supercritical potentials with $\epsilon>\epsilon_c$ correspond to the extremely thin wedge between both lines.\footnote{This wedge is actually thinner than the error in the determination of the stability boundary itself, estimated for a fixed value of $M_t$ to be $\delta M_h^\text{st}\simeq-0.5\,\text{GeV}[\alpha_s(M_Z)-0.1184]/0.0007\pm0.3\,\text{GeV}$; see \cite{Buttazzo:2013uya}.}  Comparing this wedge with the experimental $M_h-M_t$ ellipses, we see that the SM is not consistent with a theory of quantum gravity in dS space at $\sim3\sigma$.  This fact would mean that either the SM potential is compatible with a consistent dS quantum gravity theory and the experimental values of $M_h$ and $M_t$ must be very close to the stability region; or the SM potential is not compatible with a consistent dS quantum gravity theory and that fact could be used as an indication that new physics must appear below $M_P$ (modifying the shape of the effective potential so as to make it compatible with dS quantum gravity).
\begin{figure}[!t]
\centering
\resizebox{16cm}{!}{
\includegraphics{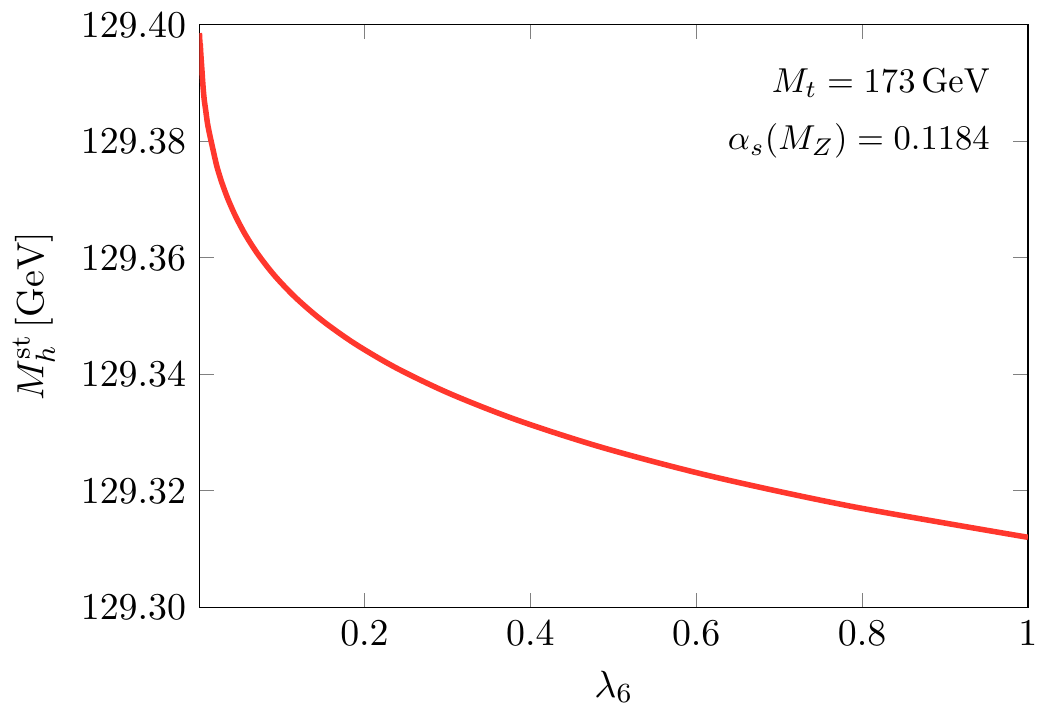}
\hspace{2cm}
\includegraphics{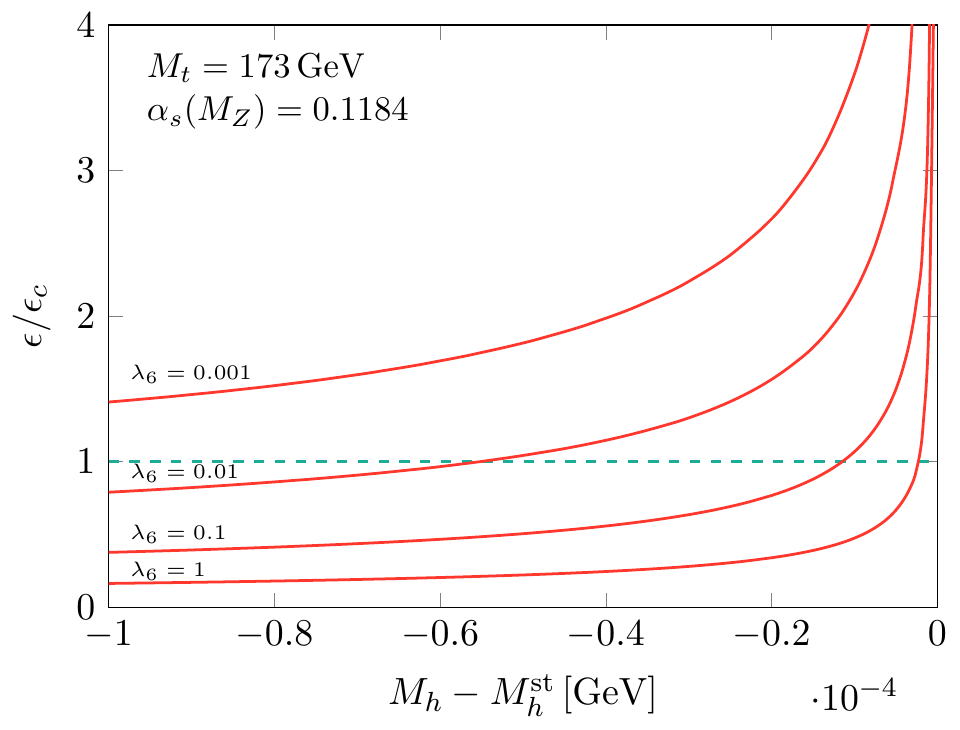}
}
\caption{Stability bound $M_h^\text{st}$ (above which the Higgs potential is stable up to $M_P$) in the presence of the term $\lambda_6h^6/(6M_P^2)$, as a function of $\lambda_6$ (left) and ratio $\epsilon/\epsilon_c$ for several values of $\lambda_6$ as a function of the difference between the Higgs mass and the stability boundary mass $M_h^\text{st}$ (right).  In both plots we have taken $M_t=173\,\text{GeV}$ and $\alpha_s(M_Z)=0.1184$.}
\label{Figlambda6}
\end{figure}

Before that last logical step is taken, however, one should also consider the possible impact of physics at the Planck scale on the shape of the potential.  This is of course difficult to achieve without knowing what this transPlanckian physics is.  One can try to capture part of the effect of that physics through non-renormalizable operators suppressed by $M_P$, with the usual limitations of effective field theories, which in our case limit explorations in field space to $h\ll M_P$.  Going beyond such limitations can only be justified as a blind parametrization of the possible impact of Planckian physics on the potential shape.  We have explored the possible impact (on the location of $\epsilon=\epsilon_c$) of adding a term $\lambda_6h^6/(6M_p^2)$ to the SM potential.  For $\lambda_6>0$ the new term adds a positive contribution to the potential and makes it more stable.  As a result, the stability boundary shown in figure \ref{FigSMStab} shifts up, towards higher values of $M_t$ that are rendered less dangerous for stability.  The shift is nevertheless very modest, as illustrated by the left plot of figure \ref{Figlambda6} that gives the Higgs mass lower bound for stability, $M_h^\text{st}$, as a function of $\lambda_6$ [for $M_t=173\,\text{GeV}$ and $\alpha_s(M_Z)=0.1184$].  The location of the line with $\alpha=1$ is also affected by the new sixtic term, as shown in the right plot of figure \ref{Figlambda6} that shows, using the thin-wall approximation, the value of $\epsilon/\epsilon_c$ for several choices of $\lambda_6$, for the same values of $M_t$ and $\alpha_s(M_Z)$.  One can see that the critical Higgs mass value, for which $\epsilon=\epsilon_c$, gets closer and closer to the stability line as $\lambda_6$ is increased, so that the width of the wedge between the critical line and the stability line is even
narrower than for $\lambda_6=0$.  From this we conclude that the effects of Planckian physics (at least those that can be captured by the naive sixtic $M_P$-suppressed term) might not modify in a significant way our previous analysis.\footnote{For completeness we also explored the less motivated case $\lambda_6<0$ together with an additional $\lambda_8h^8/(8M_p^4)$ term, with $\lambda_8>0$ (see \cite{Branchina:2014rva} and references therein).  Having $\lambda_6<0$ implies an unknown new source of potential instability from Planckian physics and generates a minimum outside the range of validity of the effective theory expansion.  Nevertheless, if we simply use it as a parametrization of the effects of Planckian physics, the stability line moves away significantly from the experimentally preferred range of $(M_h,M_t)$ strengthening our conclusions.  We have also checked that the decay rate through the new instanton solution discussed in \cite{Branchina:2014rva} is strongly suppressed when the all-important CDL gravitational effects are included.}

Therefore, if the SM is not a consistent QFT following the criteria for quantum gravity in dS space, it is necessary that new physics appears below $M_P$ for the SM to become compatible with quantum dS space.  The scale at which new physics must appear is model-dependent and cannot be determined unambiguously without committing to a particular extension of the SM.  Moreover, contrary to the second consistency condition with respect to the cosmological constant, the first consistency condition does not translate into an upper bound on the Higgs mass.  As such, it is important to point out that it does not lead to an alternative understanding of the corresponding naturalness problem.  It is unclear if there exist theories for which the first consistency condition introduced here, when used in parallel with effective potential techniques, leads to upper bounds on fundamental scalar masses.

Finally, although the constraint on the cosmological constant can only be addressed properly when $\epsilon>\epsilon_c$, it is possible to obtain a weak bound on the cosmological constant from the fact that $\xi_c\leq1/2$.  Indeed, from the SM effective potential of figure \ref{FigSMGD173} one gets that the SM vacuum energy density must satisfy $\rho_\text{vac}^{1/4}\lesssim4.7\times10^{16}\,\text{GeV}$, which is very weak when compared to the actual observed value of $\rho_\text{vac}^{1/4}=2.24\times10^{-12}\,\text{GeV}$.  Although this bound is very weak, it can be strengthened once new physics is taken into account.  Indeed, since the first consistency condition forces new physics to appear below the Planck mass and since new scalar fields lead to different inequalities relevant for the second consistency condition, it is interesting to discuss plausible extensions of the SM with new scalars.  For example, if axions \cite{Wilczek:1977pj,Weinberg:1977ma,Kim:1979if,Shifman:1979if,Zhitnitsky:1980tq,Dine:1981rt} or moduli have potentials with several minima, the second dS consistency condition brings extra inequalities for the cosmological constant.  By focusing for simplicity on QCD axions, assuming that the relevant vacuum is not the global vacuum and that the first consistency condition is satisfied, the second consistency condition leads roughly to $\rho_\text{vac}^{1/4}\lesssim\sqrt{M_aM_P}=2.7\times10^3\,\text{GeV}\sqrt{f_a/10^{10}\,\text{GeV}}$ where $M_a=0.60\,\text{meV}/(f_a/10^{10}\,\text{GeV})$ is the QCD axion mass and $f_a$ is the axion decay constant \cite{Agashe:2014kda}.  Although this constraint is still quite weak when compared to the observed value of the vacuum energy density, it is nevertheless much better than the consistency condition obtained above for the SM.  It is clear from this simple exercise that new light scalars like axions and moduli lead to stronger dS quantum gravity constraints on the cosmological constant and can help explain the hierarchy between the Planck scale and the observed cosmological constant.  It is interesting that axion scenarios can also accommodate the potential stabilization mechanism discussed in \cite{EliasMiro:2012ay}, which can move the stability boundary (and the $\alpha=1$ line with it) to be on top of the experimental values of $M_h$ and $M_t$.


\section{Discussion and Conclusion}\label{SConclusion}

We investigated in this paper the entropic implications of quantum dS spaces on scalar potentials.  Since dS space has an entropy, interpreting this entropy as the usual statistical entropy implies that a quantum theory of dS space has a finite-dimensional Hilbert space.  For a QFT with several local vacua, if the dS vacuum of interest is a global minimum, then the quantum theory of dS space is consistent.\footnote{Poincar\'e recurrences still occur but none of the type mentioned here is possible.}  If on the other hand, the dS vacuum of interest is only local and it is the minimum with the lowest positive vacuum energy density, quantum tunneling to other vacua with negative cosmological constants is possible.\footnote{This scenario is generic in supergravity.}  The quantum tunneling events, described by CDL instantons, lead to big crunches where the theory does not settle into the AdS vacua.  In such cases, for a quantum theory of dS space \textit{with a finite number of degrees of freedom} to be consistent, it is necessary to re-interpret CDL instantons entropically as recurrences to low-entropy states.  This can occur when the bounce action associated with the CDL instanton is of the order of (but smaller than) the dS entropy, leading to quantum tunneling events (almost) as often as the Poincar\'e recurrence time.  Moreover, demanding entropic consistency naturally implies two constraints on the scalar potential: the distance in field space between the dS vacuum and any other vacuum with negative cosmological constant must be of the order of the reduced Planck mass or larger; and the fourth root of the vacuum energy density of the dS vacuum must be smaller than the fourth root of the typical scale of the scalar potential.  These constraints are obtained by a careful analysis of the ratio of the instanton action to the false dS vacuum background action.  In doing so, the study of the thin-wall approximation is completed.  It is shown that there exist two qualitatively different regimes, one of them allowing an entropic interpretation of the decay.  These two regimes can be extended to generic scalar potentials, leading to the two generic constraints mentioned above.

We then argued that the two constraints should be understood as consistency conditions for QFT in dS space different in principle from the usual naturalness conditions.  Indeed, since the origin of the hierarchy problems of small scalar masses and a small cosmological constant resides in Planck scale physics, it is plausible that our understanding of physics at very high energies leads us in the wrong direction.  An unknown mechanism at the Planck scale could explain away the large additive radiative corrections and a natural theory could instead be one satisfying the two constraints introduced here.   The first constraint, with the help of the RG flow, can be translated into bounds on physical scalar masses.  The second constraint acts directly on the vacuum energy density, leading to an upper bound on the cosmological constant.  However, we stress that the constraints do not address directly the large quantum corrections to fundamental scalar mass terms and the cosmological constant.  Thus, although our constraints are exact within our assumptions (finiteness of the Hilbert space, entropic interpretation of CDL instantons), the step from constraints to consistency conditions superseding naturalness criteria is speculative.  A more complete understanding of quantum gravity (string theory comes to mind, although there exists an important no-go theorem \cite{deWit:1986xg,Maldacena:2000mw}) could shed light on the relation between the consistency conditions presented here and the usual naturalness criteria.

Finally, we worked out the implications of our constraints on the SM and its extensions.  We find that, for the SM to be dS consistent according to our assumption, the experimental values of the Higgs boson and top masses must be extremely close to the stability line.  As this seems to be disfavored at present, one can argue that new physics must appear below the Planck mass in order to make the model consistent.  This results from the first dS consistency condition, which in terms of the Higgs mass, is approximatively equivalent to the condition of stability of the SM effective potential.  Unfortunately, the constraint translates into a lower bound on the Higgs mass, forbidding a resolution of the corresponding hierarchy problem.  We analyzed the effect of new non-renormalizable operators in the SM effective potential and concluded that new physics is necessary below the Planck scale for the first dS consistency condition to be satisfied if $M_h$ and $M_t$ are too far from the stability line.  The new physics scale could unfortunately not be determined without committing to a particular extension of the SM.  We then investigated the second dS consistency condition on the cosmological constant and obtained a very weak constraint.  We also sketched how the hierarchy between the Planck scale and the observed cosmological constant can be alleviated greatly when considering plausible extensions of the SM with new light scalars, allowing a possible understanding of the smallness of the observed cosmological constant.


\ack{
JRE thanks the Physics Department of Universit\'e Libre de Bruxelles for hospitality during the final stages of this work.  JFF is pleased to thank Tom Banks for enlightening discussions.  The work of JRE is supported by the Spanish Ministry MEC under grants FPA2014-55613-P, FPA2013-44773-P, and FPA2011-25948; by the Generalitat de Catalunya grant 2014-SGR-1450; and by the Severo Ochoa excellence program of MINECO (grant SO-2012-0234).  The work of JFF and MT is supported by NSERC.
}


\bibliography{ConsistentQFTandQGdS}

\end{document}